%% file: KPark_EPJA_DVMP_Piplus.accepted.tex
\documentclass[epj]{svjour}
\usepackage{graphicx}% Include figure files
\usepackage{dcolumn}% Align table columns on decimal point
\usepackage{colordvi}
\usepackage{url}
\usepackage{longtable}
\usepackage{psfig}
\begin{document}
\hugehead 
\title{Deep exclusive $\pi^+$ electroproduction off the proton at CLAS}

\input{authors06112012}
\date{\today}

\abstract{The exclusive electroproduction of $\pi^+$ above the resonance region was studied using the $\rm{CEBAF}$ Large Acceptance Spectrometer ($\rm{CLAS}$) at 
Jefferson Laboratory by scattering a 6~GeV continuous electron beam off 
a hydrogen target. The large acceptance and good resolution
of $\rm{CLAS}$, together with the high luminosity, 
allowed us to measure the cross section for the $\gamma^* p \to n \pi^+$ process 
in 140 ($Q^2$, $x_B$, $t$) bins: $0.16<x_B<0.58$, 
$1.6$ GeV$^2<$ $Q^2$ $<4.5$ GeV$^2$ and $0.1$ GeV$^2<$ $-t$ $<5.3$ GeV$^2$. For most bins, the
statistical accuracy is on the order of a few percent.
Differential cross sections are compared to four theoretical models, 
based either on hadronic or on partonic degrees of freedom. The four models can describe the 
gross features of the data reasonably well, but differ strongly in their 
ingredients. In particular, the model based on Generalized Parton Distributions (GPDs) 
contain the interesting potential to experimentally access transversity GPDs.
} %end of abstract

\maketitle

%%%%%%%%%%%%%%%%   Introduction    %%%%%%%%%%%%%%%%%%%%%%%%%  
\section{Introduction}
\begin{figure}[!htb]
\vspace{95mm} 
\centering{\includegraphics{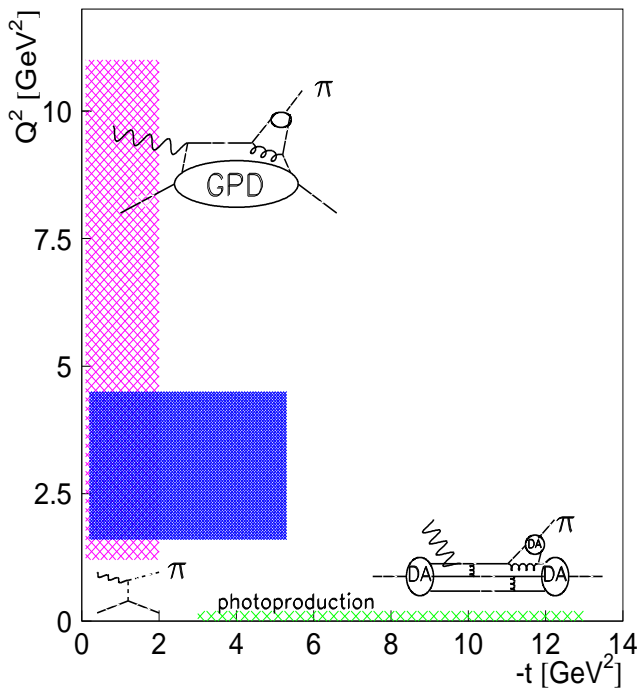}}
\caption{(color online). Schematic representation of the $\gamma^* N \to  N^\prime \pi$ process
(above the resonance region)
in different regions of the ($Q^2$, $t$) plane. The Feynman diagrams describe the reaction in terms of 
meson exchanges at low $Q^2$ and $|t|$, in terms of GPDs at large $Q^2$ and small $|t|$,
and in terms of hadron distribution amplitudes (DA) at large $|t|$. The lightly shaded areas (magenta and green online)
represent approximately the experimentally explored regions up to now. 
The darkly shaded area (blue online) represents the phase space covered by this work.}
\label{fig:motivation}
\end{figure}

One of the major challenges in contemporary nuclear physics is the
study of the transition between hadronic and partonic pictures 
of the strong interaction. At asymptotically short distances, 
the strong force is actually weak and the appropriate degrees of freedom
are the quarks and gluons (partons) whose interaction can be quantified
very precisely by perturbative Quantum Chromodynamics (pQCD).
However, at larger distances on the order of one Fermi, effective theories that take 
hadrons as elementary particles whose interactions are described
by the exchange of mesons appear more applicable.
The connection between these two domains is not well understood.
In order to make progress, a systematic study of a
series of hadronic reactions probing these intermediate distance scales 
is necessary. The exclusive electroproduction of a meson (or of a photon) 
from a nucleon,  $\gamma^* N \to N^\prime M$,
is particularly interesting. Indeed, it offers two ways to vary the scale 
of the interaction and therefore to study this transition regime. One can vary 
the virtuality of the incoming photon 
$Q^2=-(p_e-p_e^\prime)^2$, which effectively represents the transverse size of the probe, 
or the momentum transfer to the nucleon $t=(p_N-p_N^\prime)^2$, which 
effectively represents the transverse size of the target. Here, $p_e$ and
$p_e^\prime$ are the initial and scattered electron four-momenta and 
$p_N$ and $p_N^\prime$ are the initial and final nucleon four-momenta, respectively. Figure~\ref{fig:motivation} 
sketches the transition regions that have been experimentally explored 
until now (lightly shaded areas) as a function of these two variables, $Q^2$ and $|t|$. 
In this figure, we keep, quite arbitrarily, only the experiments for 
which $|t|>$ 3 GeV$^2$ in photoproduction ($\rm{SLAC}$~\cite{RAnderson00} 
and $\rm{JLab}$~\cite{WChen}) and $Q^2>$ 1.5 GeV$^2$ in electroproduction 
(Cornell~\cite{bebek76,bebek78}, $\rm{JLab}$~\cite{Horn09,HPBlok,XQian} 
and $\rm{HERMES}$~\cite{Hermes}). These are the domains for which, we believe, 
there are chances to observe first signs that partonic degrees of freedom
play a role in the reactions. The darkly shaded area 
in Fig.~\ref{fig:motivation} represents the phase space covered by the present work. It is divided 
into 140 ($Q^2$, $x_B$ or $W$, $t$) bins, to be compared to only a few  ($Q^2$, $x_B$ or $W$, $t$) bins
in the lightly shaded areas for the previous electroproduction experiments.

We also display in Fig.~\ref{fig:motivation} three Feynman-type diagrams illustrating the
mechanisms believed to be at stake for the $\gamma^* N \to N^\prime M$ process: at asymptotically large-$Q^2$,  
asymptotically large-$|t|$ (both in terms of
partonic degrees of freedom) and at low-$Q^2$ and low-$|t|$ (in terms of hadronic degrees of freedom).

At asymptotically large $Q^2$ and small $|t|$ (along the vertical axis in Fig.~\ref{fig:motivation}),
the exclusive electroproduction of a meson should be dominated by the so-called ``handbag diagram"~\cite{muller,ji,rady,JCCollins}. The initial virtual photon hits a 
quark in the nucleon and this same quark, after a single gluon exchange, 
ends up in the final meson. A QCD factorization theorem~\cite{JCCollins} states that the complex quark and gluon non-perturbative structure of the nucleon 
is described by the Generalized
Parton Distributions (GPDs). For the $\pi^+$ channel at leading twist in QCD,
 i.e. at asymptotically large $Q^2$, the longitudinal part of the 
cross section $\sigma_L$ is predicted to be dominant over the transverse part $\sigma_T$.
Precisely, ${d\sigma_L}/{dt}$ should scale as ${1}/{Q^6}$ at fixed $x_B$ and $|t|$, while 
${d\sigma_T}/{dt}$ should scale as ${1}/{Q^8}$. It is predicted that $\sigma_L$ 
is sensitive to the helicity-dependent GPDs $\tilde{E}$ and $\tilde{H}$~\cite{JCCollins}
while, if higher-twist effects are taken into account and factorization is assumed, $\sigma_T$ 
is sensitive to the transversity GPDs, $H_T$ and $\bar{E}_T=2\tilde{H}_T+E_T$~\cite{GK09}.

At large values of $|t|$, in photoproduction (i.e. along the horizontal 
axis in Fig.~\ref{fig:motivation}) but also presumably in electroproduction,
the $\gamma^{(*)} N \to N^\prime M$ process should be 
dominated by the coupling of the (virtual) photon to one of the valence 
quarks of the nucleon (or of the produced meson), with minimal 
interactions among the valence quarks. In this regime, a QCD factorization
theorem states that the complex 
structure of the hadrons is described by distribution
amplitudes (DA) which at small distances (large $|t|$)
can be reduced to the 
lowest Fock states, i.e. 3 quarks for the nucleon and 
$q$-$\bar{q}$ for the meson~{\cite{lepage}}. At sufficiently high 
energy, constituent counting rules (CCR)~\cite{SBrodsky00} 
predict an $s^{-7}$ scaling 
of the differential cross section ${d\sigma}/{dt}$ at 
fixed center-of-mass pion angles, provided $|s|$, $|t|$, 
and $|u|$ are all large. Here $s=W^2$ is the squared invariant mass
of the $\gamma^*$-$p$ system and $u=(p_\gamma^*-p_N^\prime)^2$ is given in terms
of the four-vectors $p_\gamma^* = p_e - p_e^\prime$ and $p_N^\prime$ for the final-state nucleon. 
The large $|t|$ and $|u|$ region corresponds
typically to a center-of-mass pion angle $\theta_{\rm{cm}}\approx 90^{\circ}$.
In this domain, 
the CCR predict ${d\sigma}/{dt} = f(\theta_{\rm{cm}}) s^{2-n}~$ 
for the energy dependence of the cross section, where $f(\theta_{\rm{cm}})$ depends 
on details of the dynamics of the process and $n$ is the total number of point-like 
particles and gauge fields in the initial and final states. For example, our reaction 
$\gamma^* p \to n\pi^+$ should have $n=9$, since there is one
initial photon, three quarks in the initial and the final nucleons, and
two in the final pion.

%% From which $Q^2$, from which $s$, such scaling laws start to appear 
%% are open questions. 
Many questions are open, in particular at which $Q^2$ and $s$ do such scaling laws start to appear. 
Even if these respective scaling regimes are not reached
at the present experimentally accessible $Q^2$ and $s$ values, 
can one nevertheless extract GPDs or DAs, provided that some corrections 
to the QCD leading-twist mechanisms are applied?
Only experimental data can help answer such questions.

\section{Insights from previous experiments with respect to partonic approaches} 
\label{insights}

The two most recent series of experiments that have measured exclusive $\pi^+$ electroproduction
off the proton, in the large-$Q^2$, low-$|t|$ regime 
where the GPD formalism is potentially applicable, have been conducted in $\rm{Hall~C}$ at 
Jefferson Lab ($\rm{JLab}$)~\cite{Horn09,HPBlok,XQian} and at $\rm{HERMES}$~\cite{Hermes}.

The $\rm{Hall~C}$ experiments, with 2 to 6 GeV electron beam energies,
separated the $\sigma_L$ 
and $\sigma_T$ cross sections of the $\gamma^* p \to n \pi^+$ 
process using the Rosenbluth technique for $0.17<x_B<0.48$ and  
$Q^2$ up to $3.91$ GeV$^2$.
The term $\sigma_L$ dominated the cross section
for $|t|<$ 0.2 GeV$^2$, while $\sigma_T$ was dominant for larger
$|t|$ values. These data were compared to two GPD-based 
calculations, hereafter referred to as VGG~\cite{VGG00} and GK~\cite{GK09,GK11} from the initials of the models' authors. The comparison of the data with the VGG model can be found in the
Hall C publications~\cite{Horn09,HPBlok} while the comparison with the GK model can
be found in the GK publications~\cite{GK09,GK11}.
For $\sigma_L$, which should be the QCD leading-twist contribution, these GPD calculations 
were found to be in general agreement with the magnitude and the $Q^2$- and $t$-
dependencies of the experimental data. In these two calculations 
the main contribution to $\sigma_L$ stems from the $\tilde{E}$ GPD,
which is modeled either entirely as pion-exchange in the $t$-channel~\cite{VGG00} 
or is at least dominated by it~\cite{GK09,GK11} (see Refs.~\cite{manki,franki} for
the connection between the $t$-channel pion-exchange and the $\tilde{E}$ GPD). This term is 
also called the ``pion pole", and the difference between the two calculations 
lies in the particular choice made for the $t$-channel pion propagator
(Reggeized or not) and the introduction of a hadronic form factor or not
at the $\pi NN$ vertex. In both calculations, $\sigma_L$ contains higher-twist
 effects because the pure leading-twist component of the pion pole 
largely underestimates the data. Only the GK model,
which explicitly takes into account higher-twist quark transverse momentum, is able 
to calculate $\sigma_T$. Agreement between data and calculation is found only 
if the $H_T$ transversity GPD is introduced, which makes up most of 
$\sigma_T$. 
%In summary, the normalization
%and kinematical dependencies of the separated $\sigma_L$ and $\sigma_T$ 
%cross sections of $\rm{JLab}$ $\rm{Hall~C}$ seem to be interpretable in terms 
%of GPD-based models
%if higher-twist effects, in the form of quark transverse momentum
%dependence and transversity GPDs, are taken into account.

The $\rm{HERMES}$ experiment used 27.6 GeV electron and positron beams 
to measure the $\gamma^* p \to n \pi^+$ cross section at four 
($x_B$, $Q^2$) values, with $x_B$ ranging from 0.08 
to 0.35 and $Q^2$ from 1.5 to 
 5 GeV$^2$. Since all data were taken at 
a single beam energy, no longitudinal/transverse separation could be carried
out. The differential cross section ${d\sigma}/{dt}$ was 
compared to the same two GPD models mentioned above. The GK model,
which calculates both the longitudinal and transverse parts of the cross section,
displays the same feature as for the lower energy
$\rm{JLab}$ data, i.e. a dominance of $\sigma_L$ up to $-t\approx$ 0.2 GeV$^2$,
after which $\sigma_T$ takes over. The sum of the transverse
and longitudinal parts of the cross section calculated
by the GK model is in very good agreement with the data over most of the $\-t$ range 
measured at $\rm{HERMES}$~\cite{GK09,GK11}. The VGG model, 
which calculates only the longitudinal part of the cross section, is in 
agreement with the data only for low $t$ values~\cite{Hermes}. Again,
in both calculations, $\sigma_L$ is dominated by the $\tilde{E}$ GPD,
modeled essentially by the pion pole term, and $\sigma_T$, in the GK model,
is due to the transversity GPDs. The $\rm{HERMES}$ experiment also
measured the transverse target spin asymmetry $A_{UT}$ for the 
$\gamma^* p \to n \pi^+$ process, which indicate~\cite{GK09,GK11} 
that the transversity GPDs $H_T$ or $\bar{E}_T$
indeed play an important role in the process, confirming the approach of the GK group.

The comparison between the $\rm{JLab}$ $\rm{Hall~C}$ and HERMES experiments and
the two GPD-based calculations yields very encouraging signs that, 
although higher-twist contributions definitely play a major role,
these data can be interpreted in terms of GPDs, in particular transversity GPDs. 
More precise and extensive data would be highly useful to confirm
these findings. Firstly, the present CLAS experiment extends somewhat the 
($x_B$, $Q^2$) phase space previously covered by the $\rm{JLab}$ $\rm{Hall~C}$ experiments 
and secondly, it covers 20 ($x_B$, $Q^2$) bins (with statistical errors of a few percent 
on average) which doubles the number of bins of the $\rm{JLab}$ $\rm{Hall~C}$ experiments 
(and triples the $\rm{HERMES}$ number of bins).
These new data are important to test the present GPD-based model 
calculations and, if successful, bring more stringent constraints 
on the current GPD parametrizations.

The large-$|t|$ (large-$|u|$) domain, where the DA formalism is asymptotically 
applicable for $\gamma^{(*)} p \to n \pi^+$, has so far  been explored 
only in high-energy photoproduction at $\rm{SLAC}$~\cite{RAnderson00} and 
intermediate-energy photoproduction at $\rm{JLab}$~\cite{LYZhu}. 
While the $\rm{SLAC}$ data tend to follow the $s^{-7}$ scaling asymptotic prediction,
for a $90^{\circ}$ center-of-mass angle, 
the more recent $\rm{JLab}$ data, which are compatible with the $\rm{SLAC}$ data
but are more precise, actually reveal some large oscillations around this $s^{-7}$ behavior.

In recent years a similar trend, i.e. ``global" scaling behavior, has been 
observed in deuteron photo-disintegration experiments~\cite{JNapolitano00,CBochna00,ESchulte00,PRossi00},
and also in hyperon photoproduction~\cite{Schumacher:2010qx}.
It would be interesting to see this in exclusive pion electroproduction
and if so, whether the oscillations disappear as $Q^2$ increases. 
The measurement presented in this article is the first one to explore
this large-$|t|$, large-$|u|$ domain ($\theta_{\rm{cm}}\approx 90^{\circ}$) 
for $\sqrt{s}>$ 2 GeV in $\pi^+$ exclusive electroproduction off the proton.
The present CLAS electroproduction experiment covers a $t$-range
up to $\approx$ 5 GeV$^2$ while the largest $|t|$-values 
measured by Hall C are $\approx$ 0.9 GeV$^2$ and by HERMES $\approx$ 2 GeV$^2$.

\section{The Experiment} 
\begin{figure}[!htb]
\begin{center}
\includegraphics[angle=0, height=8cm,width=0.47\textwidth]{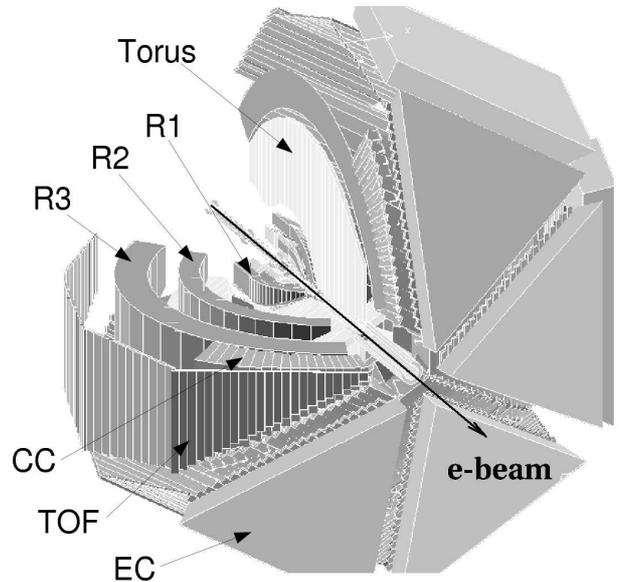}
        \caption{
          Three-dimensional view of the $\rm{CLAS}$ detector.
%Three-dimensional view of the CLAS detector, showing the three regions of drift chambers (R1, R2, R3) Cherenkov counters (CC), the time-of-flight system (TOF), and the electromagnetic calorimeter (EC) (see text for details). In this picture, the electron beam travels from the upper-left corner to the lower-right corner
          \label{fig:clas}
        }
\end{center}
\end{figure}
The measurement was carried out with the CEBAF Large Acceptance Spectrometer ($\rm{CLAS}$)~\cite{CLAS}. A schematic view of $\rm{CLAS}$ is shown in Fig.~\ref{fig:clas}. $\rm{CLAS}$ has a toroidal magnetic field generated by six flat superconducting coils (main torus), arranged symmetrically around the azimuth. Six identical sectors are independently 
instrumented with 34 layers of drift cells for particle tracking ($\rm{R1}$, $\rm{R2}$, $\rm{R3}$), plastic scintillation counters for time-of-flight ($\rm{TOF}$) measurements, gas threshold Cherenkov counters ($\rm{CC}$) for electron and pion separation, 
and electromagnetic calorimeters ($\rm{EC}$) for photon and neutron detection. 
To aid in electron/pion separation, the $\rm{EC}$ is segmented into an inner part closer to
the target and an outer part further away from the target. $\rm{CLAS}$ covers on average 80\% of the full $4\pi$ solid angle for the detection of charged particles. The azimuthal acceptance is maximum at a polar angle of $90^\circ$ and decreases at forward angles. The polar angle coverage ranges from about $8^{\circ}$ to $140^{\circ}$ for the detection of $\pi^+$. The scattered electrons are detected in the $\rm{CC}$ and $\rm{EC}$, which extend from $8^{\circ}$ to $45^{\circ}$.

 The target is surrounded by a small toroidal magnet (mini-torus). This magnet is used to shield the drift chambers closest to the target from the intense low-energy electron background resulting from M\o ller scattering. 
 
 A Faraday cup, composed of 4000 kg of lead and 75 radiation lengths thick,
 is located in the beam dump, $\approx$ 29 meters downstream the CLAS target.
 It completely stops the electrons and thus allows to measure the 
 accumulated charge of the incident beam and therefore the total
 flux of the beam~\cite{CLAS}.

 The specific experimental data set ``e1-6'' used for this analysis was collected in 2001. The incident beam had an average intensity of $7\;\rm{nA}$ and an energy of $5.754\;\rm{GeV}$. The 5-$\rm{cm}$-long liquid-hydrogen target was located $4\;\rm{cm}$ upstream of the $\rm{CLAS}$ center. This
 offset of the target position was found to optimize the acceptance of forward-going
 positively charged
 particles. The main torus magnet was set to 90\% of its maximum field, corresponding to an
 integral magnetic field of $\approx$ 2.2 Tm in the forward direction. The
 torus current during the run was very stable ($<0.03\%$). Empty-target runs were performed 
 to measure contributions from the target cell windows.

 In this analysis, the scattered electron and the produced $\pi^+$ were detected and the 
 final state neutron determined from missing mass. The continuous electron beam provided by CEBAF is well suited for measurements involving two or more final-state particles in coincidence, leading to very small accidental coincidence contributions, smaller than $10^{-3}$, for the instantaneous luminosity of $10^{34}$ cm$^{-2}$s$^{-1}$ of the present measurement.   
 
 Raw data were subjected to the calibration and reconstruction procedures that are part of the standard $\rm{CLAS}$ data analysis sequence. Stringent kinematic cuts were applied to select events with one electron candidate and only one positively charged track. These events were then subjected to further selection criteria described in the following Section. Throughout the analysis, 
the experimental data distributions were compared to the output
of our Monte Carlo program $\rm{GSIM}$ (see Sec.~\ref{data_analysis}).

\begin{figure}[!htb]
\vspace{60mm} 
\centering{\includegraphics{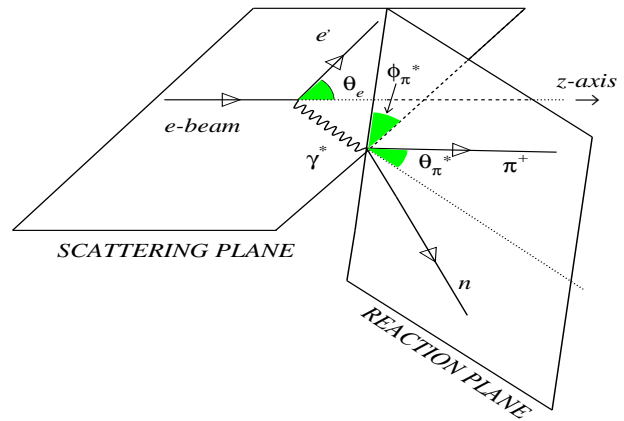}}
\caption{Kinematics of exclusive single $\pi^+$ electroproduction from a proton target.}
\label{fig:kinematics}
\end{figure}

 A schematic illustration of electron scattering off a nucleon target producing an outgoing nucleon and one pion is shown in Fig.~\ref{fig:kinematics}. The scattered electron angle $\theta_e$ is given in the laboratory frame. The angle between the virtual photon three-momentum and the direction of the pion is denoted as
  $\theta_{\pi}^*$ and the angle between the electron scattering plane and hadronic production plane
  is denoted as $\phi_{\pi}^*$. These two angles are defined in the center-of-mass frame of the hadronic system. The angle $\phi_{\pi}^*$ is defined so that the scattered electron lies in the $\phi_{\pi}^*=0^{\circ}$ half plane with the $z$-axis pointing along the virtual photon momentum. For exclusive single $\pi^+$ production from the proton, 
  we request the simultaneous detection of one single electron and of one single $\pi^+$ in CLAS
  and the final state neutron will be identified by the missing mass squared
  $((p_e + p_N)-(p_e^\prime+p_{\pi}))^2$, 
 where $p_{\pi}$ is the four-momentum of the detected $\pi^+$.  The kinematic range and
 bin sizes are chosen to provide reasonable statistics in each bin.
 These are summarized in Table~\ref{tab:kine_range}.
\begin{table}[htb]
\begin{center}
\caption{Kinematic bins used in this analysis.}
\begin{tabular}{ccccc}
\hline
Variable & Number of bins & Range & & Bin size \\
\hline
\hline
$x_{B}$  & 7  & $0.16$ - $0.58$ & & $0.06$ \\\\

$Q^2$ &  5 & $1.6$ - $3.1$ $\rm{GeV^2}$& & $0.3$ $\rm{GeV^2}$ \\
      &  3 & $3.1$ - $4.5$ $\rm{GeV^2}$& & $0.5$  $\rm{GeV^2}$\\\\

$-t$ &   6 & $0.1$ - $1.9$ $\rm{GeV^2}$& & $0.3$ $\rm{GeV^2}$\\ 
     &   3 & $1.9$ - $4.3$ $\rm{GeV^2}$& & $0.8$ $\rm{GeV^2}$\\ 
     &   1 & $4.3$ - $5.3$ $\rm{GeV^2}$& & $1.0$ $\rm{GeV^2}$\\ 
\hline
\end{tabular}
\label{tab:kine_range}
\end{center}
\end{table}

Our aim is to extract the three-fold differential cross section
$\frac{1}{\Gamma}\frac{d^3\sigma}{dQ^2dx_Bdt}$ where:

\begin{eqnarray}\label{eq:sig}
\frac{1}{\Gamma}\frac{d^3\sigma}{dQ^2dx_Bdt}=
\frac{n_{w}(Q^2, x_B, t)}{\mathcal{L}_{int} \ \Delta Q^2 \ \Delta x_B \ \Delta t},
\end{eqnarray}

\noindent with
\begin{itemize}
\item $n_{w} (Q^2, x_B, t)$ is the weighted number of 
	$ep \rightarrow e^\prime n\pi^+$ events in a given bin ($Q^2$, $x_B$, $t$);
	in particular, $n_{w}$ ($x_{B}$, $Q^2$, $-t$) contains the detector's acceptance
	correction factor $Acc$($x_{B}$, $Q^2$, $-t$, $\phi^*_{\pi}$) (see Sec.~\ref{Acceptance_Corrections})
	and the correction factor due to radiative effects $F_{rad}$($x_{B}$, $Q^2$, $-t$)
	(see Sec.~\ref{Radiative_Corrections}),
\item $\mathcal{L}_{int}$ is the effective integrated luminosity,
\item $\Delta Q^2$, $\Delta x_B$ and $\Delta t$ are the corresponding bin widths 
	(see Table~\ref{tab:kine_range}); 
	for bins not completely filled, because of $W$ or $E^\prime$ cuts on 
	the electron for instance (see fig.~\ref{fig:kinebin}), 
	the phase space $\Delta Q^2 \Delta x_B  \Delta t$
	includes a volume correction and the $Q^2$ and $x_B$ central
	values are modified accordingly.
\end{itemize}
In the following three sections, we detail the various
cuts and correcting factors entering the definition of $n_{w} (Q^2, x_B, t)$.
%
%
%

%%%%%%%%%%%%%%%%   Data Calibrations    %%%%%%%%%%%%%%%%%%%%  
\section{Data Analysis}\label{data_analysis}

%%%%%%%%%%%%%%%%%%   Data Corrections    %%%%%%%%%%%%%%%%%%%  
\subsection{Particle identification and event selection}\label{PIDselection}
% The $e p \to e\prime n\pi^+$ reaction is identified by detecting the scattered electron in %coincidence with a $\pi^+$ and by using the missing mass technique to ensure the exclusivity of %the reaction. Good identification of the electron and pion is therefore the most important %issue for the channel identification.

\subsubsection{Electron identification}
 The electrons are identified at the trigger level by requiring 
 at least 640 MeV energy deposited in the $\rm{EC}$ in coincidence with a 
 signal in the $\rm{CC}$ (which triggers on one photoelectron).
 
\begin{figure}[!htb]
\begin{center}
  \includegraphics[angle=0,height=6.5cm,width=0.4\textwidth]{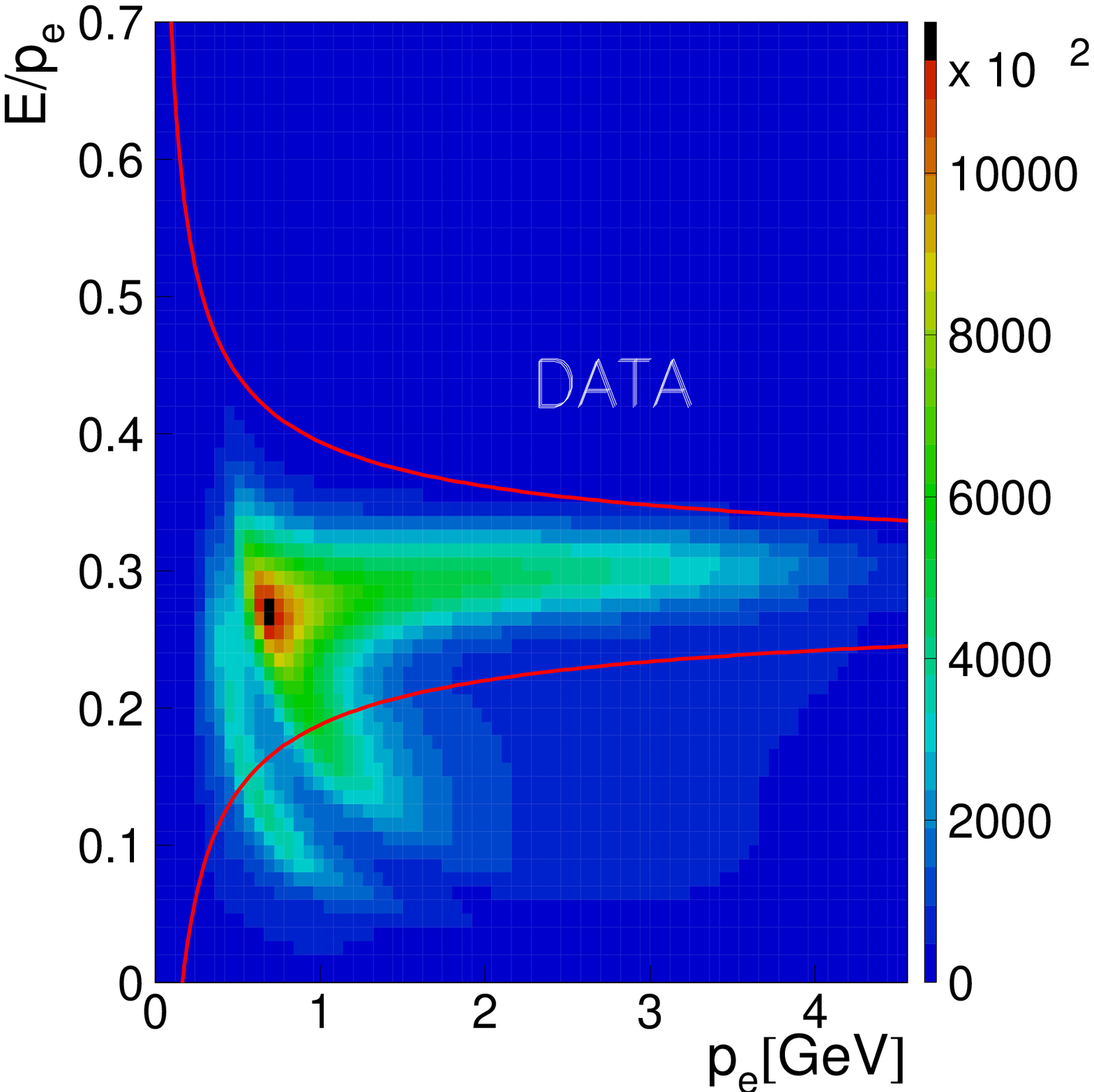}
  \includegraphics[angle=0,height=6.5cm,width=0.4\textwidth]{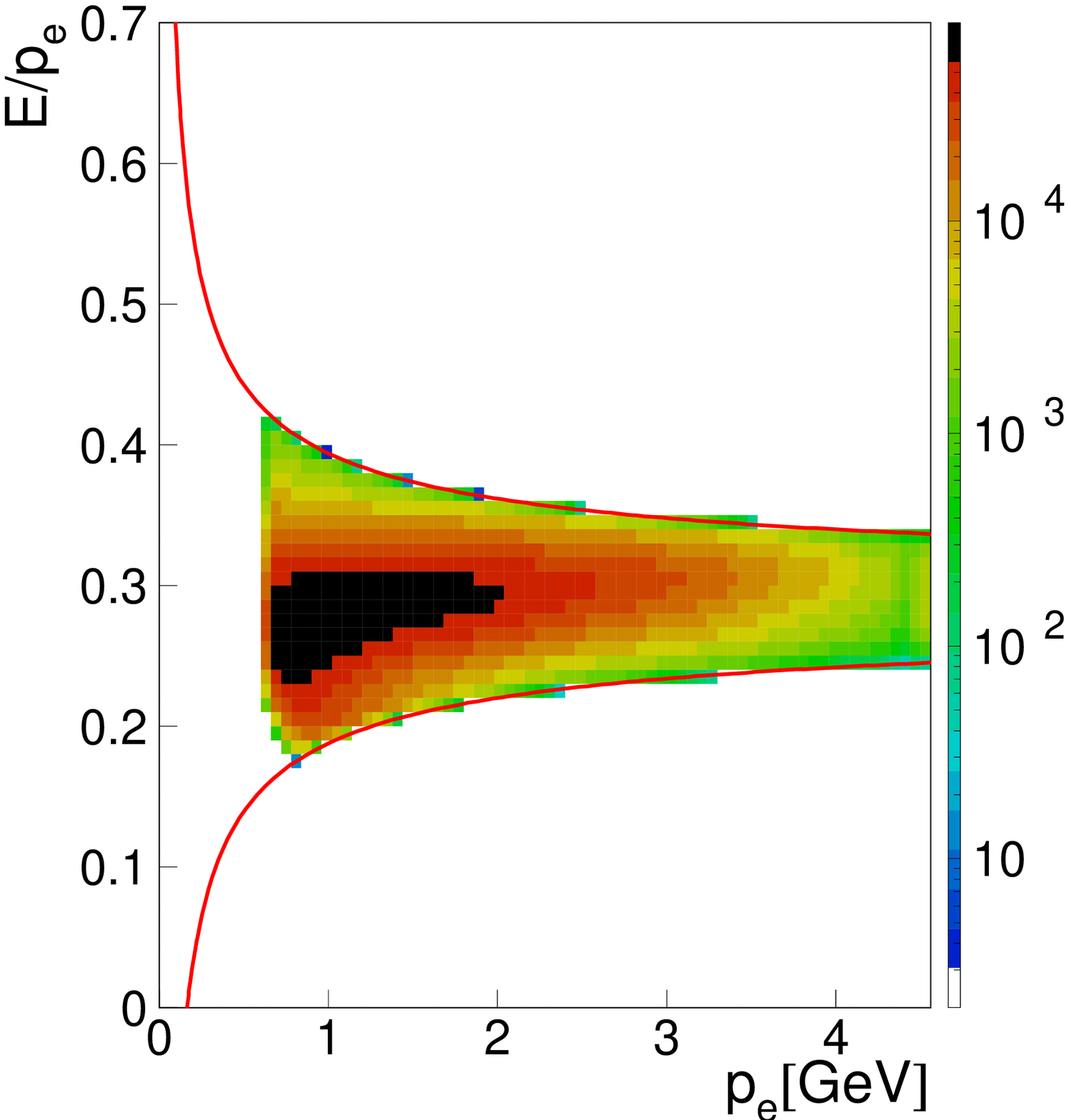}
  \caption{
    (color online). $\rm{EC}$ sampling fraction versus particle momentum for the
    experimental data before (top) and after (bottom) EC energy cuts. The solid curves show the $\pm3\sigma$ sampling fraction cuts which are applied to select electrons.
    \label{fig:ec_sf}
  }
\end{center}
\end{figure}

 Additional requirements for particle identification (PID) were used in the 
 off-line analysis to refine the electron identification. First, we required 
 that the $\rm{EC}$ and $\rm{CC}$ hits matched with a reconstructed track 
 in the drift chambers ($\rm{DC}$). Second, we correlated the energy deposited 
 in the $\rm{EC}$ and the momentum obtained by the track reconstruction in the $\rm{DC}$.
 This is aimed at removing the pion contamination. Electrons deposit energy in proportion 
 to their
 incident energy in the calorimeter whereas pions are minimum ionizing and deposit 
 a constant amount of energy in the calorimeter. 
 The ratio of the total deposited energy in the $\rm{EC}$ to
 the momentum of the particle is called the sampling fraction. For electrons, approximately 
 30\% of the total energy deposited in the $\rm{EC}$ is directly measured 
 in the active scintillator material. The remainder of the energy is 
 deposited in the lead sheets interleaved between the scintillators. 
 Figure~\ref{fig:ec_sf} shows the sampling fraction ${E}/{p_e}$ versus
 particle momentum $p_e$. The average 
 sampling fraction for 
 electrons was found to be 0.291 for this experiment. The solid lines 
 in Fig.~\ref{fig:ec_sf} show the $\pm$3$\sigma$ sampling fraction cuts used in this analysis.

To further reject pions, we required the energy deposited 
in the inner $\rm{EC}$ to be larger than $50$ MeV.
Minimum ionizing particles 
lose less than this amount in the 15 cm thickness of the inner $\rm{EC}$.

 Fiducial cuts were applied to exclude the $\rm{EC}$ detector edges.
  When an electron hit is close to an edge,
 part of the shower leaks outside the device; in this
case, the energy cannot be fully reconstructed from the
calorimeter information alone. This problem can be
avoided by selecting only those electrons lying inside a
fiducial volume within the $\rm{EC}$ that excludes the
edges. A $\rm{GEANT}$-based simulation ($\rm{GSIM}$) was used to
determine the $\rm{EC}$-response with full electron 
energy reconstruction. The calorimeter fiducial volume was
defined by cuts that excluded the inefficient detector 
regions.

 Particle tracks were reconstructed using the drift chamber information, 
 and each event was extrapolated to the target center to obtain a 
 vertex location. We demanded that the reconstructed $z$-vertex position 
 (distance along the beam axis from the center of $\rm{CLAS}$, with negative 
 values indicating upstream of the $\rm{CLAS}$ center) lies in the range 
 $-80\;\rm{mm}< z_{\rm{vtx}} < -8\;\rm{mm}$. This is slightly larger than the 
 target cell size in order to take into account the resolution effects
 on the vertex reconstruction.

 Finally, a lower limit on the number of photoelectrons detected in the photomultiplier tubes of the $\rm{CC}$ provided an additional cut to improve electron identification. The number of photoelectrons detected in the $\rm{CC}$ follows a Poisson distribution modified for irregularities in light collection efficiency for the individual elements of the array. For this experiment, a good electron event was required to have 3 or more photoelectrons detected in the $\rm{CC}$. The efficiency of the $\rm{CC}$ cut was determined from the experimental data. We fit the number of photoelectrons using the modified Poisson distribution. The efficiency range after the $\rm{CC}$ cut is 78\% to 99\% depending on the kinematic region. The correction is then the integral below the cut divided by the total integral of the resulting fit function.

\subsubsection{Positively charged pion identification}
The main cuts to select the $\pi^+$ are based on charge, $z$-vertex, fiducial cuts 
and velocity versus momentum correlations. The velocity $\beta$ is calculated from 
the ratio of the path length of the reconstructed track, to the time of flight. 

\begin{figure}[!htb]
\begin{center}
\includegraphics[angle=0,width=0.39\textwidth]{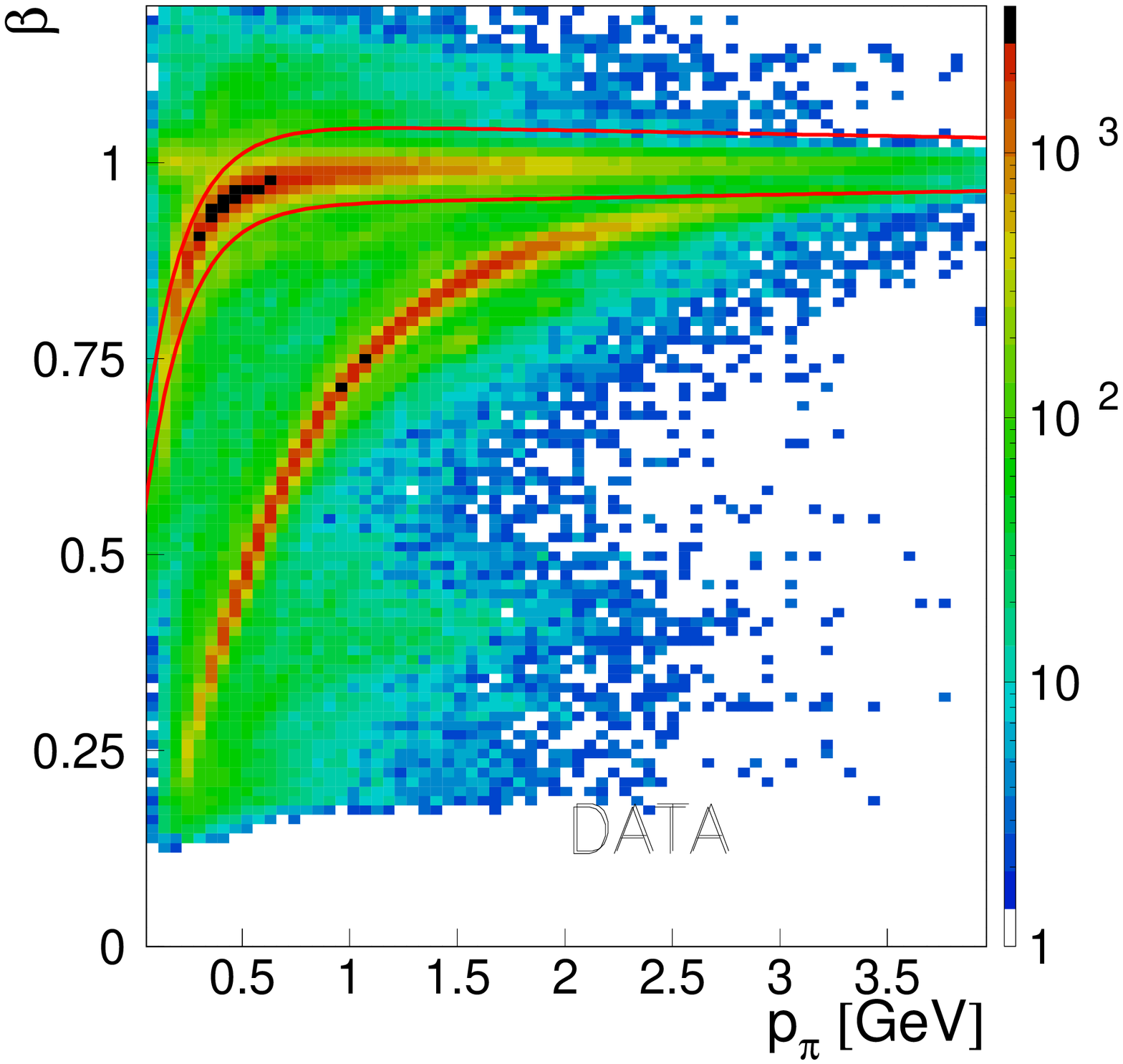}
\includegraphics[angle=0,width=0.39\textwidth]{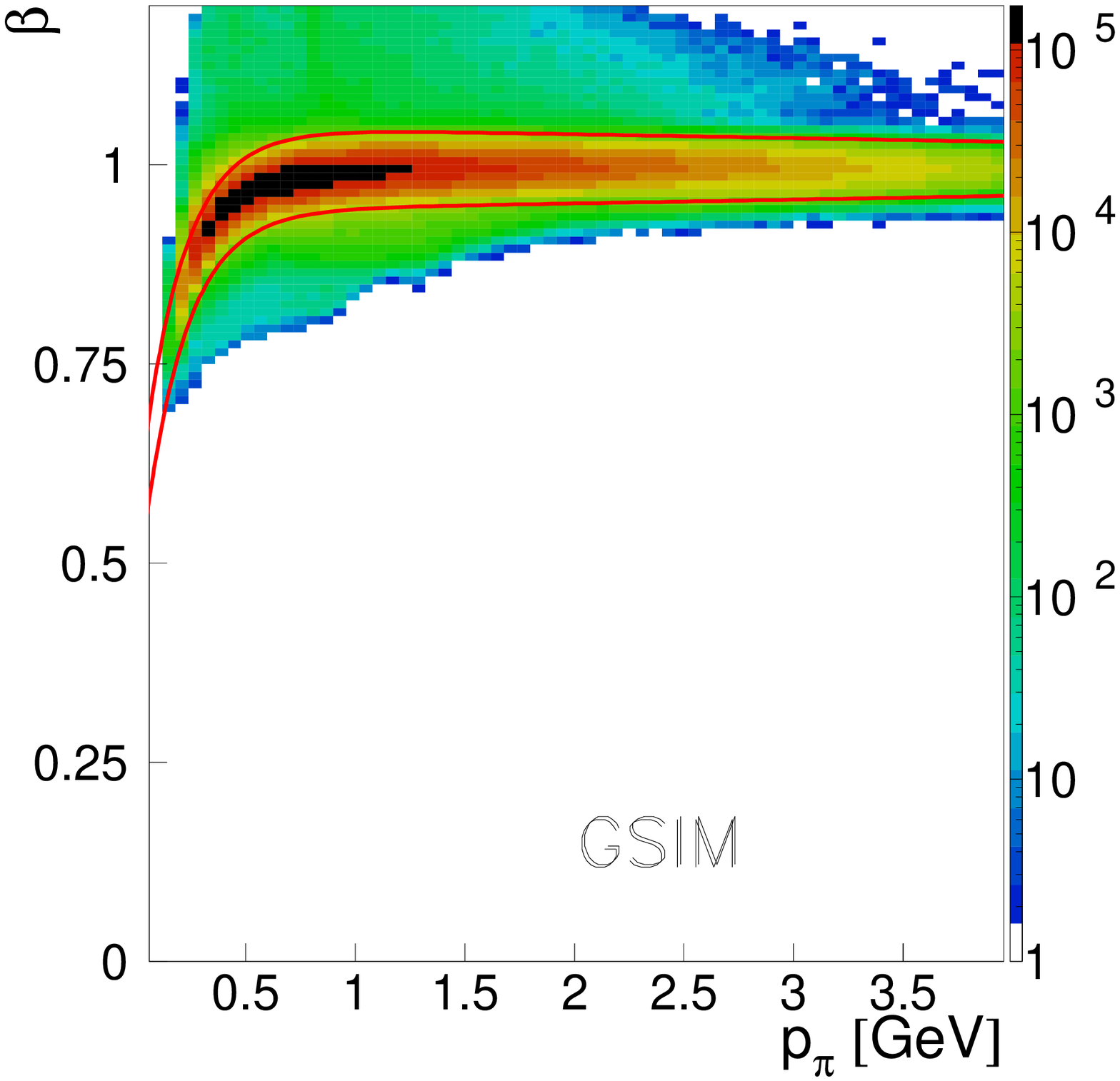}
        \caption{
          (color online).  Velocity $\beta$ versus momentum for $\pi^+$ candidates
	  using experimental data (top) and the $\rm{GSIM}$ Monte Carlo simulation (bottom). The solid 
	  curves are $\pm1.5\sigma$ $\beta$ cut lines used to select $\pi^+$ from
	  positron ($\beta$=1 band) and proton ($\beta <$0.8 band) backgrounds.
	  (The data are issued from a skimmed file which selected one electron 
and at least one positively charged particle and one neutral particle).
          \label{fig:pion_id_beta_mean}
        }
\end{center}
\end{figure}

Figure~\ref{fig:pion_id_beta_mean} shows the $\beta$ versus $p$ distribution for positively charged particles from experimental data (top) and from the $\rm{GSIM}$ Monte Carlo simulation (bottom). 
A Gaussian is fit to $\beta$ for bins in momentum $p_\pi$. A $\pm1.5\sigma$ cut on $\beta$ is chosen for pion candidates as shown in Fig.~\ref{fig:pion_id_beta_mean} (solid curves in the plot). Pions and positrons ($\beta=1$) are well separated below $p_\pi = 250\;\rm{MeV/c}$ of momentum in the experimental data, but this is no longer the case at momenta larger than $400\;\rm{MeV/c}$. For this reason, positrons 
can be mis-identified as pions, which increases the background. At higher momenta, there can also 
be some particle mis-identification from protons and kaons. We estimated
that the missing mass and vertex cuts reduce this mis-identification
to the 5 - 10\% level. This residual background contamination was subtracted 
as described in Sec.~\ref{Background_Subtraction}.

%%%%%%%%%%%%%%%% Fiducial cuts
\subsection{Fiducial cuts}\label{fiduXXX}

\subsubsection{Electron fiducial cuts}
\begin{figure}[!htb]
\begin{center}
\includegraphics[angle=0,width=0.38\textwidth]{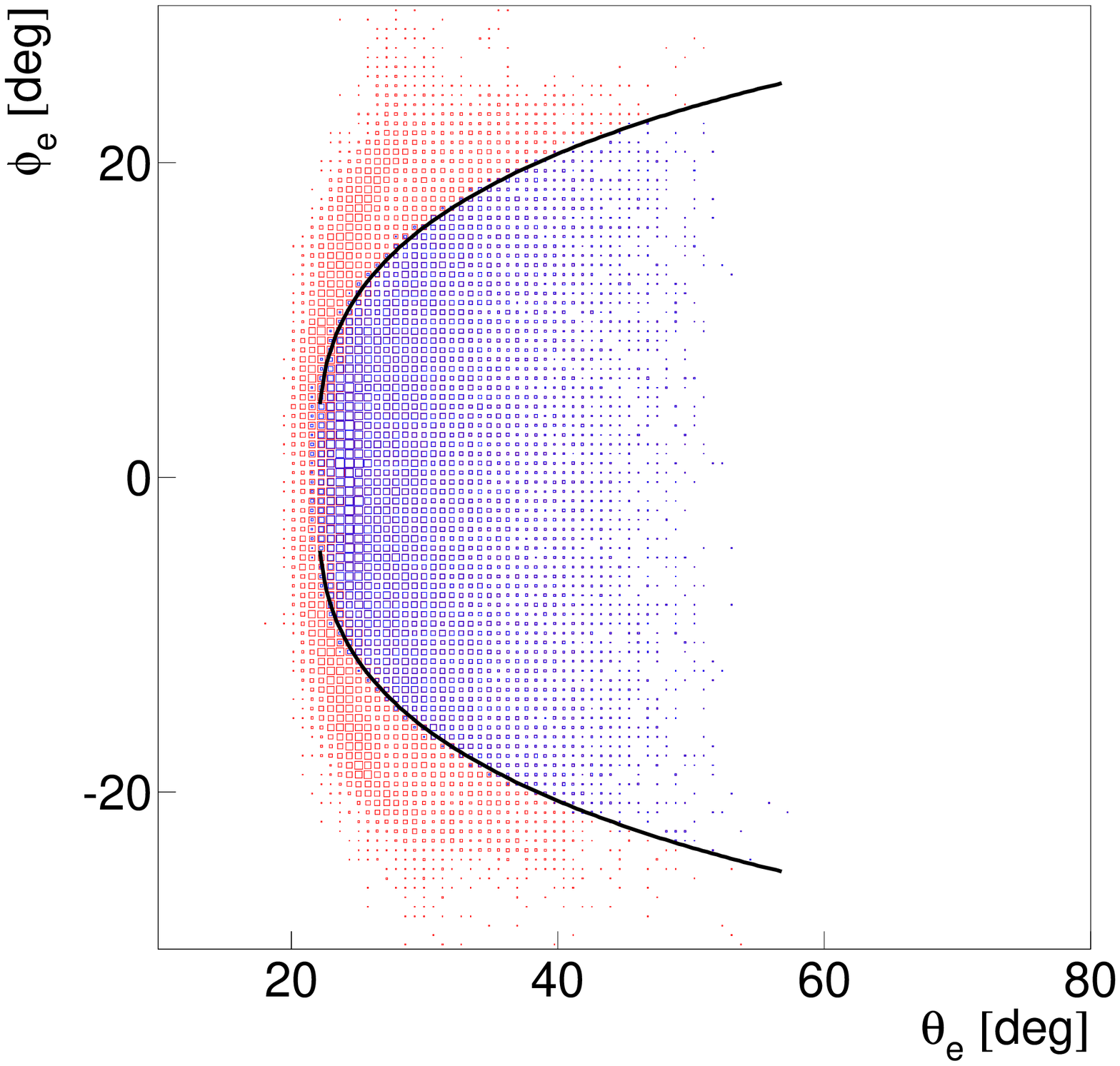}
\includegraphics[angle=0,width=0.38\textwidth]{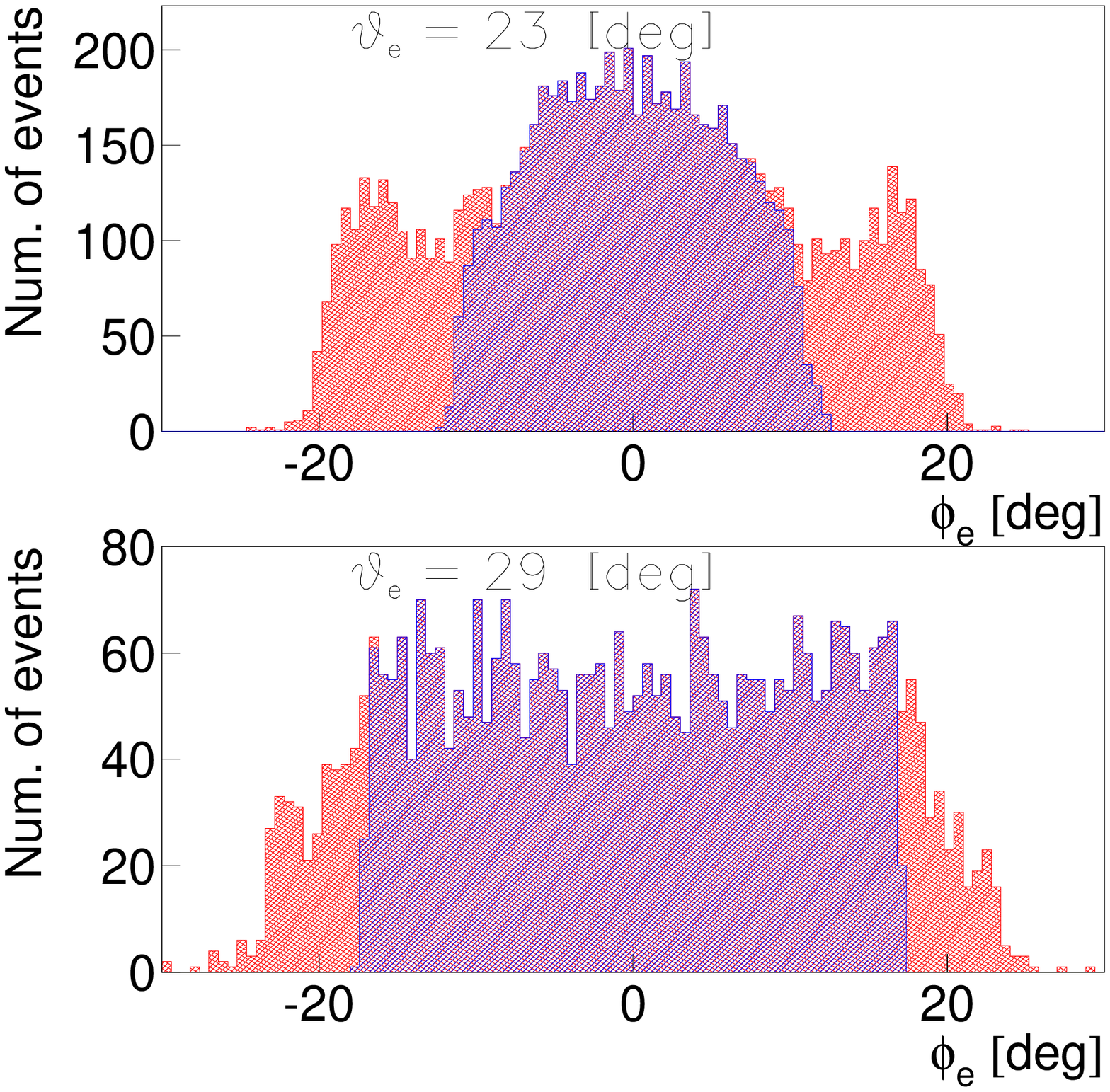}
        \caption{
          (color online). Example of electron fiducial cuts for the electron momentum bin ($p_e=1.437\;\pm.025\;\rm{GeV}$) in Sector 2. See the detailed explanation in the text.
          \label{fig:e_fcut_loose00}
        }
\end{center}
\end{figure}

 The fiducial cuts for electrons were developed to exclude regions with non-uniform detector efficiency such as the edges of a sector in the $\rm{CC}$ and $\rm{EC}$. The fiducial cut is a function of the angles $\theta_e$, $\phi_e$, and momentum $p_e$ of the electron. 
 An example of such fiducial cut can be seen in Fig.~\ref{fig:e_fcut_loose00} for a 
 given electron momentum bin. The solid line in the top plot shows the boundary of 
 the fiducial region for the central momentum in that bin. Only electron events inside 
 the curve (blue area) were used in the analysis. This curve was determined by selecting 
 the flat high-efficiency areas in the $\theta_e$-sliced $\phi_e$ distributions. The histograms 
 on the bottom of Fig.~\ref{fig:e_fcut_loose00} show examples of such $\phi_e$ distributions 
 at two values of $\theta_e=23^{\circ}\pm 0.5^{\circ}$ and $29^{\circ}\pm0.5^{\circ}$.
 One sees a central, uniform area, flanked by two fringes.
 The highlighted area in the center indicates the selected fiducial range.
 In addition, a set of $\theta_e$ versus $p_e$ cuts was used to eliminate the areas 
 with low detection efficiency due to problematic time-of-flight counters, photomultiplier 
 tubes in Cherenkov counters, or drift chamber areas. 
%
%
%%%%%%%%%%%%%%%%%%%%%%%%%%%%%%%%%%%%%%%%%%%%%%%%%%%%%%%%%%%%%%%%%%%%%%%%%%%%%%%%%%%%%
%/data/scratch/parkkj/ana_note_scaling/review_plot/fidus

\subsubsection{Pion fiducial cuts}\label{pion_fidu}
\begin{figure}[htb]
\vspace{3mm}
\begin{center}
        \includegraphics[angle=0,width=0.4\textwidth]{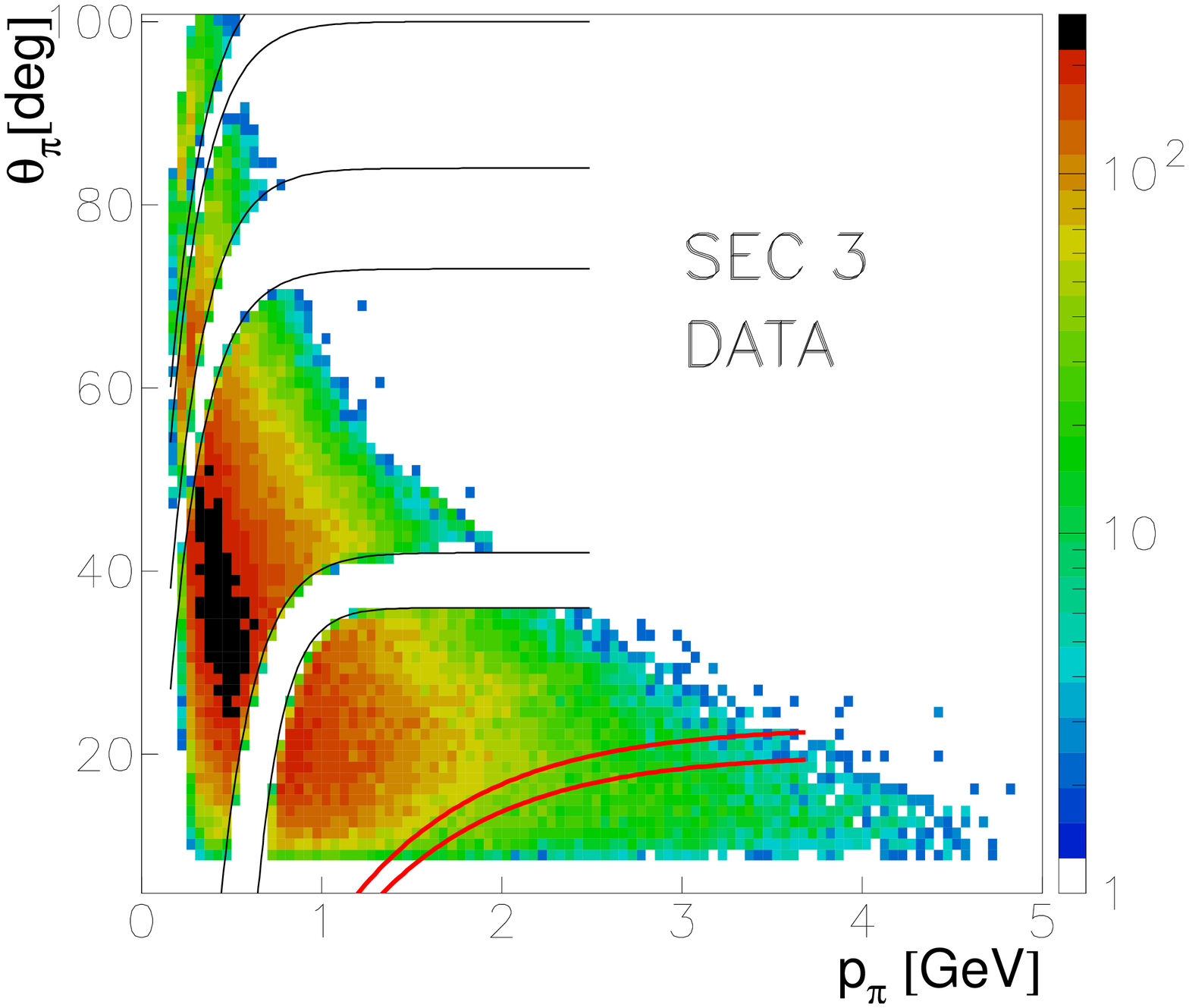}
        \includegraphics[angle=0,width=0.4\textwidth]{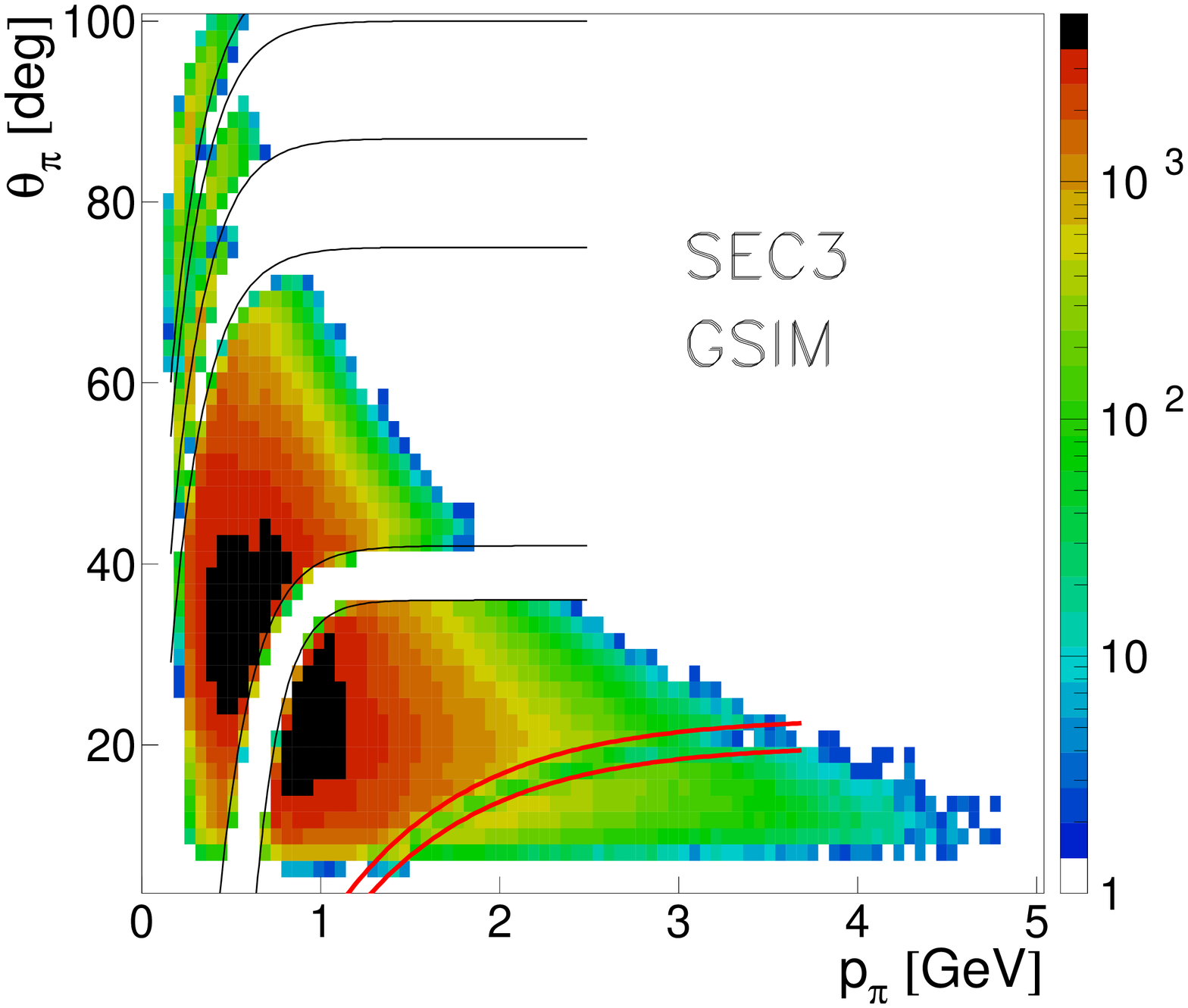}
        \caption{
           (color online).
          \label{fig:pvstheta} Pion polar angular distribution as a function of momentum in Sector 3. The low detector response areas are removed by empirical cuts for experimental (top) and simulated data (bottom). The black thin solid curves are fiducial cuts based on $\rm{DC}$ inefficiencies and the red thick solid curves are cuts for bad $\rm{TOF}$ counters. 
        }
\end{center}
\end{figure}

 The fiducial cuts for pions depend on the angles $\theta_{\pi}$, $\phi_{\pi}$ and the momentum $p_{\pi}$. The pion momentum is scanned in $100\;\rm{MeV}$ steps from $0.3$ to $1.7\;\rm{GeV}$. The uniform detector efficiency region was determined by selecting a flat high-efficiency $\phi_{\pi}$ region in each $\theta_{\pi}$-sliced momentum bin, and the bad $\rm{TOF}$ counters and the inefficient $\rm{DC}$ areas were excluded by additional software 
 cuts (the same procedure as was applied to electrons). Figure~\ref{fig:pvstheta} shows an example for the fiducial cuts for pions. The low-efficiency $\rm{DC}$ regions (between the black solid lines) and the bad $\rm{TOF}$ paddles (between red solid lines) are removed 
 in both experimental (top) and simulated (bottom) data as part of the fiducial cuts. 
%/data/scratch/parkkj/ana_note_scaling/review_plot/fidus/appfidu2011pion.kumac
%/data/scratch/parkkj/ana_note_scaling/review_plot/gsimplot02.prc.kumac

\subsection{Kinematic corrections}\label{Kinematics_Corrections}

Due to effects that are not included in the 
reconstruction software (deviations of the magnetic field from perfect toroidal symmetry,
misalignment of the tracking system,...), we have to apply some empirical corrections 
to the measured angles and momenta of both electrons and pions. For electrons, the kinematic
corrections are applied using the elastic $ep\to e^\prime p^\prime$ process for which the kinematics 
is over-constrained. The goal is to correct the three-momentum of the electron so as to
minimize the constraints due to the equations of conservation of energy and momentum.
The same procedure is applied to the $\pi^+$'s three-momentum using our reaction  
$ep \to e^\prime \pi^+n$ under study, minimizing the deviation of
the missing mass peak position from the neutron mass. The same correction factors are 
used for all events having the same kinematics. In this way we keep the spatial resolution 
of the drift chamber systems and multiple scattering effects and the missing mass resolution 
approaches its intrinsic limitations. The corrections were  
most sizable ($\approx$ 5\%) for the pion momentum. They resulted in an improved missing mass resolution, 
from 23 to 35 MeV depending on kinematics. The corrections were most sizable for the 
high-momentum and forward-angle pions at high $W$ which are of
interest in this experiment. 
We then applied additional ad-hoc smearing factors for the tracking and timing resolutions to the Monte Carlo so that they match the experimental data.

\section{Monte Carlo simulation}\label{Monte_Calro_Simulation_Data}
\begin{figure}[htb]
\begin{center}
        \includegraphics[angle=0,width=0.38\textwidth]{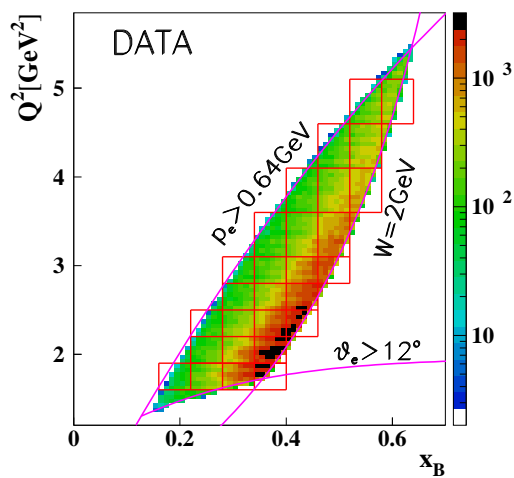}
        \caption{
           (color online).
          \label{fig:kinebin} Kinematic coverage and binning (red boxes) as a function of $x_{B}$ and $Q^2$ (integrated over all other variables) for the experimental data.  
	  Only events with $W > 2\;\rm{GeV}$ are shown.
        }
\end{center}
\end{figure}

 In order to calculate the CLAS acceptance for $ep \to e^\prime \pi^+n$, 
 we simulated electron and pion tracks using 
 the CLAS $\rm{GEANT3}$-based Monte Carlo Package $\rm{GSIM}$. 
 For systematic checks, we used two Monte Carlo event generators.
 Our approach is that by comparing the results of simulations
 carried out with two very different event generators, a 
 conservative and reliable estimation of systematic effects,
 such as finite bin size effects, is obtained.  
 The first event generator, $\rm{GENEV}$ (see Ref.~\cite{genev} for 
 the original publication dedicated to photoproduction processes), 
 generates events for various
 exclusive meson electroproduction reactions for proton and neutron targets
  ($\pi$, $\omega$, $\rho^0$, and $\phi$), including their decay, radiative effects, and resonant and non-resonant multi-pion production, with realistic kinematic distributions.
 $\rm{GENEV}$ uses cross 
 section tables based on existing photoproduction data and extrapolates to electroproduction
 by introducing a virtual photon flux factor ($\Gamma$) and the electromagnetic form factors.
 Radiative effects, based on the Mo and Tsai formula~\cite{MoTsai00}, 
 are part of this event generator as an option. Although the
 formula is exact only for elastic $e$-$p$ scattering, it can be used
 as a first approximation to simulate the radiative tail and to 
 estimate bin migration effects in our pion production process, as will be discussed in Sec.~\ref{Radiative_Corrections}.
 The second event generator, 
 $\rm{FSGEN}$~\cite{fsgen}, distributes events according to 
 the $ep \to e^\prime \pi^+n$ phase space.\\ 
  Electrons and positive
 pions were generated under the ``e1-6'' experimental conditions. 
Events were processed through $\rm{GSIM}$.
  As already mentioned, additional ad-hoc smearing factors for the tracking and timing resolutions 
  are applied after $\rm{GSIM}$ so that they match 
the experimental data. The low-efficiency regions in the drift chambers and problematic $\rm{TOF}$ channels were removed during this procedure. Acceptance and radiative corrections were
 calculated for the same kinematic bins as were used for the yield 
 extraction as 
 shown in Table~\ref{tab:kine_range}. Figure~\ref{fig:kinebin} shows the
  binning in $Q^2$ and $x_{B}$ applied in this analysis. However, some bins will be dropped 
  at some later stage in the analysis, in particular due to
very low acceptances (see following subsection).
 Our cross sections will be defined at the ($x_{B}, Q^2, -t$) values 
given by the geometrical center of the three-dimensional bins.
To account for non-linear variations of the cross section within a bin, a correction to our cross sections is determined by fitting with a simple ad-hoc three-variable function the simultaneous ($x_{B}, Q^2, -t$)-dependence 
of our cross sections. This correction comes out at the level of a couple of percent in average.  
%/data/scratch/parkkj/ana_note_scaling/review_plot/kine

\subsection{Acceptance corrections}\label{Acceptance_Corrections}
 We related the experimental yields to the cross sections using the acceptance, including the efficiency of the detector.  The acceptance factor ($Acc$) compensates for various effects, such as the geometric coverage of the detector, hardware and software inefficiencies, and resolution 
 from track reconstruction. We generated approximately $850$ million events, taking radiative effects into account. This results in a statistical uncertainty for the acceptance determination of less than 5\%
 for most bins, which is much lower than the systematic uncertainty that 
we have estimated (see Sec.~\ref{syst}).
  
\begin{figure}[htb]
\vspace{3mm}
\begin{center}
	\includegraphics[angle=0, height=8cm, width=0.45\textwidth]{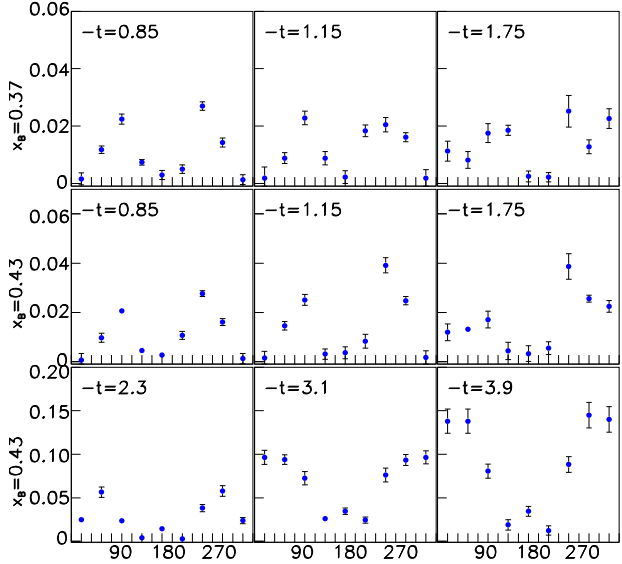}
        \caption{
          (color online). Examples of acceptance as a function of $\phi_{\pi}^*$ for various $t$ and $x_{B} $ bins at $Q^2 = 2.35\;\rm{GeV}^2$. The 
	   dips at $\phi_{\pi}^*$ = 0$^\circ$ and 180$^\circ$ are due 
	  to sector boundaries in $\rm{CLAS}$.
          \label{fig:acc00}
        }
\end{center}
\end{figure}

 We define the acceptance as a function of the kinematic variables,

\begin{eqnarray}\label{eq:ratio2}
Acc(x_{B}, Q^2, -t, \phi^*_{\pi}) =  \frac{N^{REC}(x_{B}, Q^2, -t, \phi^*_{\pi})}{N^{GEN}_{rad.}(x_{B}, Q^2, -t, \phi^*_{\pi})}~,
\end{eqnarray}
where $N^{REC}$ is the number of reconstructed particles and $N^{GEN}_{rad.}$ is the number of 
generated particles in each kinematic bin (the meaning of the subscript $rad.$ will
become clear in the next section). The kinematic variables 
in $N^{GEN}$ refer to the generated values so that
bin migration effects are taken into account in the definition of our acceptance. The acceptances are
in general between 1 and  9\%. Figure~\ref{fig:acc00} shows examples of acceptances, determined with the $\rm{GENEV}$+$\rm{GSIM}$ packages, as a function of the angle $\phi_{\pi}^*$ at a given $Q^2$ for 
various $x_{B}$ and $t$ bins. Bins with an acceptance below 0.2\% were dropped.
For the 
integration over the $\phi_{\pi}^*$ angle, in order to obtain our three-fold cross sections,
we fitted the acceptance-corrected $\phi_{\pi}^*$ distributions, so that any hole in the $\phi_{\pi}^*$ distribution
would be replaced by its fit value.

% acc = 0.097
%
% DIR : /data/scratch/parkkj/ana_note_scaling/review_recal_crs/gpp_gsim

\subsection{Radiative correction}\label{Radiative_Corrections}
 
Our goal is to extract the so-called Born cross section (tree-level) of the 
$p(e,e^{\prime}\pi^+)n$ which can thus be compared to models. However, we measure a
process which is accompanied by higher order radiative effects. Our experimental
cross section must therefore be corrected. Radiative corrections are of two
types: ``virtual" corrections where there is no change in the final state of the
$p(e,e^{\prime}\pi^+)n$ process and ``real" ones where there is in addition one (or several)
Bremsstrahlung photon(s) in the final state. Such real Bremsstrahlung photons can originate 
either from the primary hard scattering at the level of the target proton (internal radiation) 
or from the interaction of the scattered or the initial electron with the various
material layers of the $\rm{CLAS}$ detector that it crosses (external radiation).

\begin{figure}[htb]
\vspace{5mm}
\begin{center}
        \includegraphics[angle=0,height=5cm,width=0.47\textwidth]{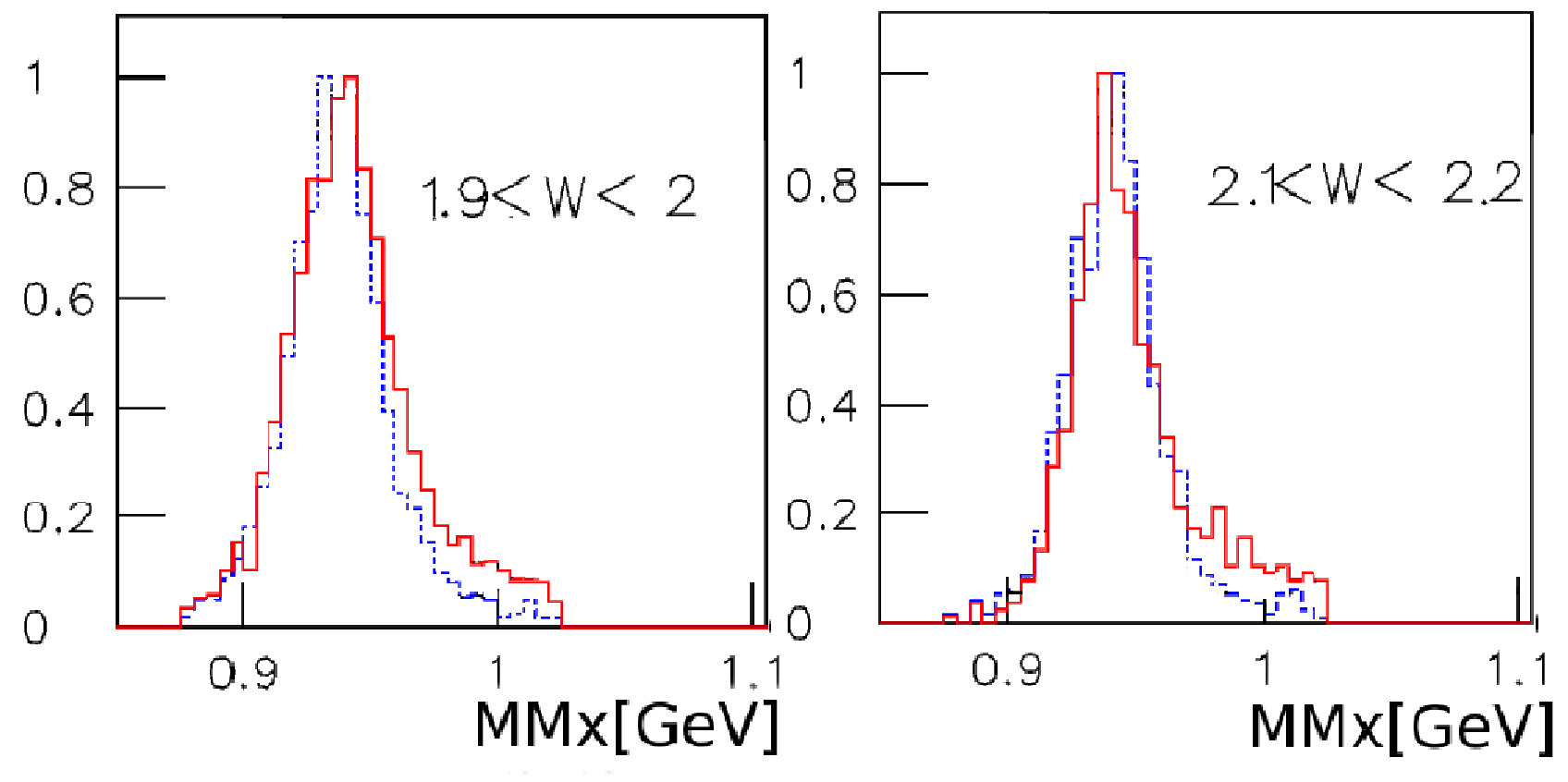}
        \caption{(color online). The simulated neutron missing mass distributions for two
	$W$ bins with $\Delta W =100\;\rm{MeV}$ at $W=1.95$ (left) and $2.15$ (right)
	$\;\rm{GeV}$ integrated over $\phi_{\pi}^*$, $\cos\theta_\pi^*$, and $Q^2$. 
	Normalized yields are shown with (solid red) and without (dashed blue)
	radiative effects.
          }
          \label{fig:frad00}
\end{center}
\end{figure}

We have dealt with these corrections in two steps.
The effects of the radiation of hard photons 
 (for instance, the loss of events due to the application of 
 a cut on the neutron missing mass) are taken into account by the
 Monte Carlo acceptance calculation described in the previous section.
 Indeed, as mentioned earlier, the $\rm{GENEV}$ code has the option to generate radiative photons
 according to the Mo and Tsai formula and the $N^{GEN}_{rad.}$ events in Eq.~\ref{eq:ratio2}
 were generated with this option turned on.
 Figure~\ref{fig:frad00} shows examples of the simulated neutron missing 
 mass with and without radiative effects in two $W$ bins, obtained 
 with the $\rm{GENEV}$ event generator and $\rm{GSIM}$. Again, the Monte Carlo simulations were 
 carried out with the same cut procedures and conditions as used in the data analysis.

Then, the correction due to soft photons and virtual corrections
 is determined by extracting the ratio between the number of events 
 without radiative and with radiative effects at the level of $\rm{GENEV}$
  for each three-dimensional kinematic bin. We therefore apply the following 
  additional correction factor to our data:
  
\begin{eqnarray}\label{eq:frad}
   F_{rad}(x_{B}, Q^2, -t)=\frac{N^{GEN}_{no rad.}(x_{B}, Q^2, -t)}
  {N^{GEN}_{rad.}(x_{B}, Q^2, -t)}.
 \end{eqnarray}

\begin{figure}[htb]
\vspace{5mm}
\begin{center}
        \includegraphics[angle=0,width=0.3\textwidth]{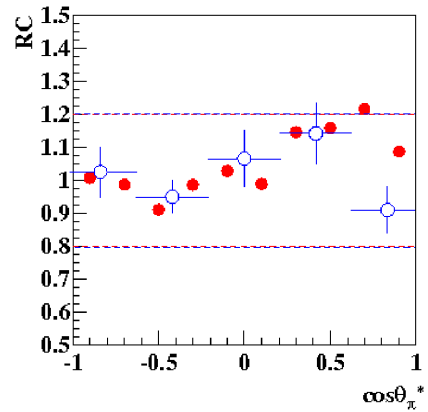}
        \caption{
          (color online). Radiative-correction factors (RC) as a function of $\cos\theta_\pi^*$ from $\rm{EXCLURAD}$ (red solid points) at $W=1.74\;\rm{GeV}$, $Q^2=3\;\rm{GeV^2}$, and $\phi_\pi^* = 112.5^{\circ}$ and $\rm{GENEV}$ plus $\rm{GSIM}$ 
	  (blue open circle) at $W \approx 1.75\;\rm{GeV}$, $Q^2 \approx 3\;\rm{GeV}^2$, and  $80^{\circ} < \phi_\pi^* < 120^{\circ}$.}
          \label{fig:rc_calculation2}
\end{center}
\end{figure}

 As a check, these radiative-correction factors were also calculated with 
the $\rm{EXCLURAD}$ code~\cite{afanasev},
 which contains a complete description of all internal radiative effects in exclusive processes,
but is currently 
 valid only up to $W=2\;\rm{GeV}$. We compare the two different radiative-correction methods in a kinematic region where both methods are valid.  Figure~\ref{fig:rc_calculation2} shows the results for radiative-correction factors in the region $W \approx 1.75\;\rm{GeV}$ and $Q^2 \approx 3\;\rm{GeV}^2$ as a function of $\cos\theta^*_{\pi}$. 

 The radiative correction factors from $\rm{EXCLURAD}$ are within $\pm20\%$ of
 unity over the full  $\cos\theta_\pi^*$ range (red solid points). The radiative corrections from $\rm{GENEV}$+$\rm{GSIM}$ also fluctuate  around $1.0$ with a similar structure (blue open circles). The $\rm{GENEV}$+$\rm{GSIM}$ error bars 
 are due to Monte Carlo statistics ($\rm{EXCLURAD}$ is a theoretical code which has therefore no
 statistical uncertainty).
 The agreement between the two approaches is important because $\rm{EXCLURAD}$ is believed
to be the most reliable of the two methods because it does not have
the limitations of Mo and Tsai. Building on this
 reasonable agreement in this part of the phase space, we rely on the $\rm{GENEV}$+$\rm{GSIM}$ 
 radiative-correction factors for our data. In Sec.~\ref{syst}, we discuss 
the systematic uncertainty associated with these radiative corrections.

\section{Background subtraction}\label{Background_Subtraction}

There are two main sources of background in our reaction. One consists of 
the mis-identification of pions with other positively charged 
particles (protons, kaons, positrons). This is particularly important 
for the pion-proton separation at high-momenta ($p>2\;\rm{GeV}$), see Sec.~\ref{PIDselection}. The other consists of multi-pion 
production. To subtract both backgrounds, we fit the neutron missing 
mass distribution bin by bin. We used many methods to fit these
spectra: fit of only the background, fit of the signal plus background,
with different functional forms both for the signal and the background,
variation of the fitted range, etc... from which we extracted
a systematic uncertainty (see Sec.~\ref{syst}).

\begin{figure}[htb]
\begin{center}
        \includegraphics[angle=0,width=0.33\textwidth]{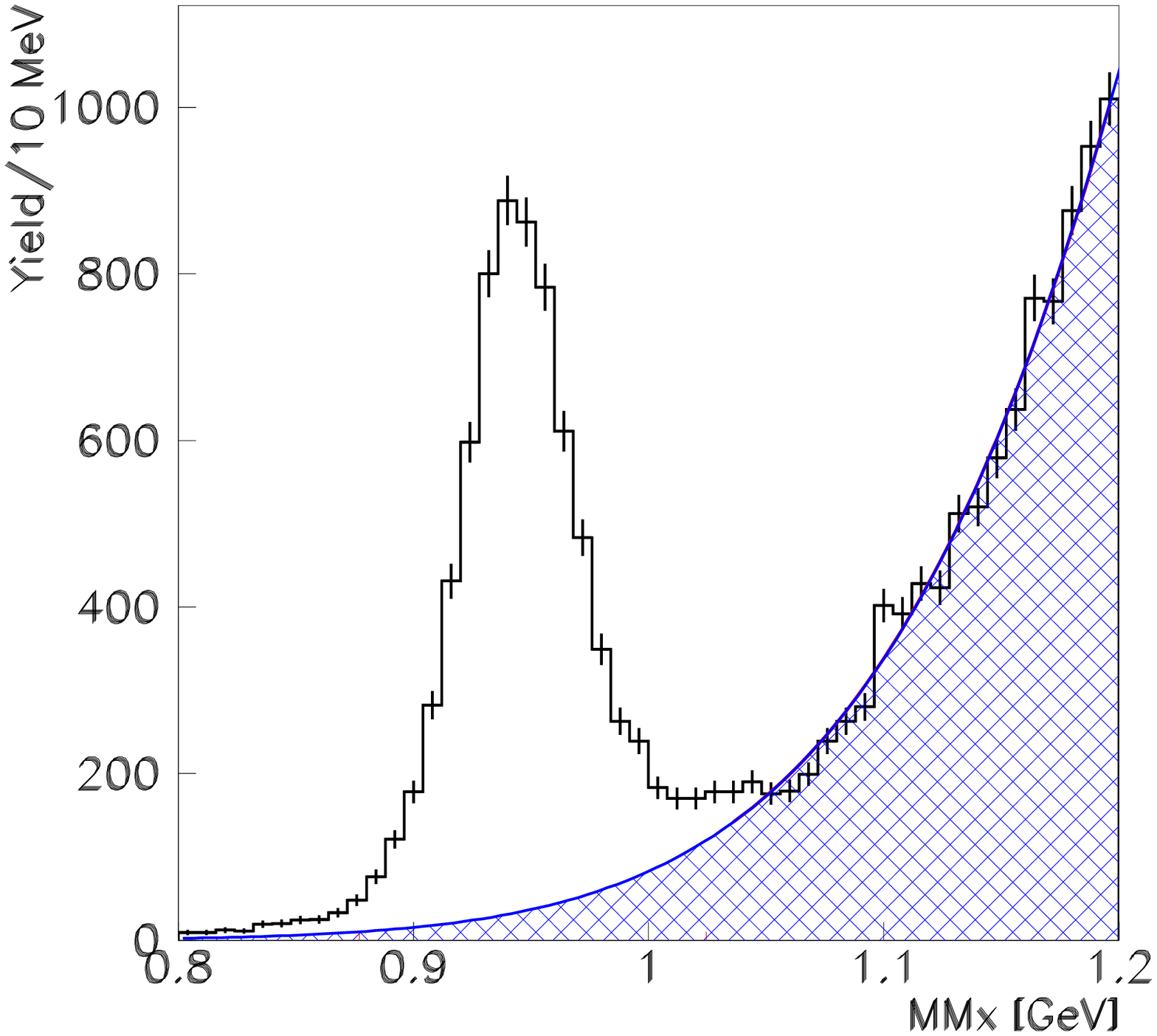}
        \includegraphics[angle=0,width=0.33\textwidth]{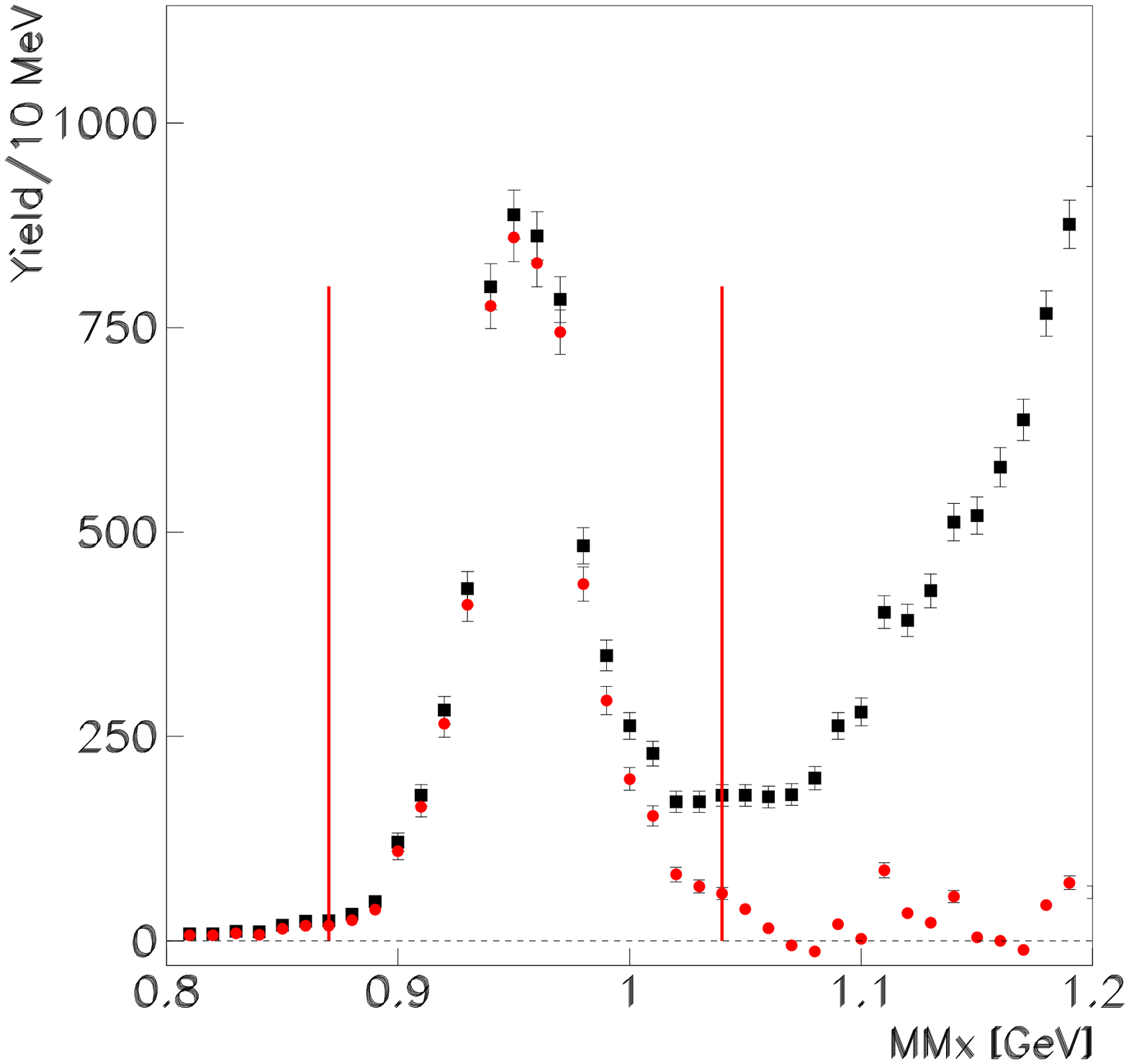}
        \caption{(color online). Example of the missing mass peak plus background at $Q^2 = 2.65\;\rm{GeV}^2$, $-t = 1.15\;\rm{GeV^2}$, and  $x_{B}=0.43$. The top plot shows the fitted background distribution (hashed region). The bottom plot shows the neutron missing mass distribution before (black squares) and after (red solid points) background subtraction. 
          }
          \label{fig:bg_subt2}
\end{center}
\end{figure}

 Figure~\ref{fig:bg_subt2} (top) shows an example of a fit based on
only the background, with an exponential 
plus a Gaussian. The former function was determined from simulations 
of the multi-pion spectra in the neutron missing mass region 
$> 1.02\;\rm{GeV}$.
 A comparison of the missing mass ($\rm{MMx}$) spectrum is shown in the bottom plot of 
 Fig.~\ref{fig:bg_subt2} before (black squares) and after (red solid circles) 
 background subtraction. In the range of the neutron missing mass cut,
 shown by the two vertical lines at $0.877\;\rm{GeV}$ and 
 $1.0245\;\rm{GeV}$, the background is small, and the remaining radiative 
 tail becomes visible after the background is subtracted.

% pion-proton mis-ID : analysis note Fig.36
%## file : /data/scratch/parkkj/ana_note_scaling/review_plot_fig25_26

\section{Systematic uncertainties}
\label{syst}

Several sources of systematic uncertainty that can affect our 
measurements have been studied by changing various cuts and using 
different event generators. 

We varied the criteria used for the particle identification to provide 
more and less stringent particle selection simultaneously for experimental 
and GSIM data and then reran the complete analysis.
 The cuts on $\rm{EC}$ energy deposition and $\rm{CC}$ amplitude for
the electron, as well as cuts on the $\rm{TOF}$ timing for the pion, have been varied.
The $\rm{EC}$ sampling fraction cut was varied from $\pm3\sigma_{\rm{EC}}$ to $\pm2\sigma_{\rm{EC}}$ which 
 led to a 5\% uncertainty for electron identification.
 Changing the $\rm{TOF}$ $\beta$ cut from $\pm2\sigma_{\rm{TOF}}$ to $\pm2.5\sigma_{\rm{TOF}}$ 
 for pion identification gives a 1.7\% uncertainty.
 The various cuts for channel identification 
such as fiducial, missing mass, and vertex cuts produced 3\%, 1\%, and 
1.6\% systematic uncertainties, respectively. 

Acceptance and radiative corrections are the biggest sources of 
systematic errors. The systematic uncertainty from the acceptance 
is evaluated by comparing our results using the $\rm{GENEV}$ and $\rm{FSGEN}$ event 
generators. In the limit of infinitely large statistics and infinitely 
small bin size, our acceptances should be model-independent (up to the bin-migration effects). 
But these conditions are not reached 
 here and we find differences between 
2 and 8\%. The systematic uncertainty for radiative 
corrections is estimated similarly by comparing the radiative-correction 
factors ($\rm{GENEV}$ and $\rm{EXCLURAD}$). We calculated the difference between
the cross sections corrected for radiative effects using either $\rm{GENEV}$-$\rm{GEANT}$ simulation or 
the $W$-expanded $\rm{EXCLURAD}$ (where $\rm{EXCLURAD}$ was linearly extrapolated to $W>2$ GeV). An average 
$8\%$ systematic uncertainty was found. Acceptance and radiative corrections
are actually correlated, but after a combined analysis we estimated an averaged range $4-12\%$ total 
uncertainty for both of these effects together. 

Concerning the background subtraction procedure
under the neutron missing mass (see Sec.~\ref{Background_Subtraction}),
we used various fitting functions (Gaussian plus exponential, 
Gaussian plus polynomial, exponential plus polynomial, etc.) and various fitting ranges.
These various fitting functions and ranges eventually produced small 
differences and we estimated a 3$\%$ systematic uncertainty
associated with this procedure.

To take into account the model-dependency of our bin-centering correction (see Sec.~\ref{Kinematics_Corrections}),
we also introduce an error equal to the correction factor itself which is, we recall, 
at the level of a couple of percent in average.

These latter systematic uncertainties were determined for each bin.
Concerning overall scale uncertainties, the target length and density 
have a 1\% systematic uncertainty and the integrated charge
uncertainty is estimated at 2\%. The background from the target cell was subtracted based on
the empty-target runs and amounted to 0.6$\pm$0.2\% of our $e^{\prime}\pi^+X$ events. 
The total systematic uncertainties, 
averaged over all bins, is then approximately 12\%.
 Table~\ref{tab:sys} summarizes the main systematic uncertainties in this analysis averaged over 
all the accessible kinematic bins seen in Fig.~\ref{fig:kinebin}.

\begin{table} [!htb]
\begin{center}
\caption{Average systematic uncertainties for the differential cross sections.}
\vspace{2mm}
\begin{tabular}{lcr}
\hline
Source & Criterion   & Estimated\\
 &    &contribution\\
\hline \hline
Type & point-to-point   & \\
\hline
$e^-$ PID & sampling fraction  & \\
 &  cut in $\rm{EC}$ & \\
  &  ($3 \sigma_{\rm{SF}}\to 2 \sigma_{\rm{SF}}$)  &  { 5\%} \\\\

$e^-$ fiducial cut  & fiducial volume change & \\
                    & ($10\%$ reduced) & {2.5\%} \\\\

${\pi}^+$ PID & $\beta$ resolution change& \\
  & ($2 \sigma_{\rm{TOF}}\to 2.5 \sigma_{\rm{TOF}}$) &  {1.7\%} \\\\

${\pi}^+$ fiducial cut &  width ($10\%$ reduced) &  {3.5\%} \\\\

Missing & neutron missing & \\
 mass &  mass resolution & \\
cut   &($3 \sigma_{\rm{MMx}} \to 3.5 \sigma_{\rm{MMx}}$)  & {1\%} \\\\

Vertex cut & $z$-vertex width  &       \\
           & ($5\%$ reduced)     & {1.6\%}\\\\

Acceptance & $\rm{GENEV}$ vs $\rm{FSGEN}$ &   \\
%% Radiative & $\rm{GENEV}$ vs $\rm{EXCLURAD}$ &  {9.5\%(3-12\%)} \\
Radiative & $\rm{GENEV}$ vs $\rm{EXCLURAD}$ &  {4-12\%} \\
corrections & &   \\\\

Background  &  various fit functions    &   \\
subtraction &  exponential, gaussian    &\\
            &and high order polynomials & 3\%\\\\
Bin-centering & toy model &  {2-4\%} \\
effect & &   \\\\
\hline
\hline
Type & overall scale/normalization   & \\
\hline
LH2 target & density/length &  {1\%} \\\\

Luminosity  & integrated charge  &  {2\%} \\\\
\hline\hline

Total  &   &  {9-14\%} \\
\hline
\end{tabular}
\label{tab:sys}
\end{center}
\end{table}

\vspace{3mm}
\section{Results and Discussion}
 In this section, we present our results for the cross sections of the $p(e,e^{\prime}\pi^+)n$ reaction in the invariant mass region $W>2\;\rm{GeV}$.  We have extracted the differential cross sections as a function of several variables ($t$, $Q^2$, and $W$ or $x_B$).
 The angle $\phi^*_{\pi}$ is always integrated over in the following. 
 The extraction of the interference cross sections $\sigma_{TT}$
and $\sigma_{TL}$ is the subject of an ongoing analysis and will be 
presented in a future article. 
The error bars on all cross sections include both statistical and systematic uncertainties added in quadrature.
All values of our cross sections and uncertainties can be found on 
the CLAS database web page: http://clasweb.jlab.org/cgi-bin/clasdb/db.cgi

\subsection{${d\sigma}/{dt}$ as a function of $t$}

\begin{figure*}[!htb]
\vspace{210mm}
\centering{
\includegraphics{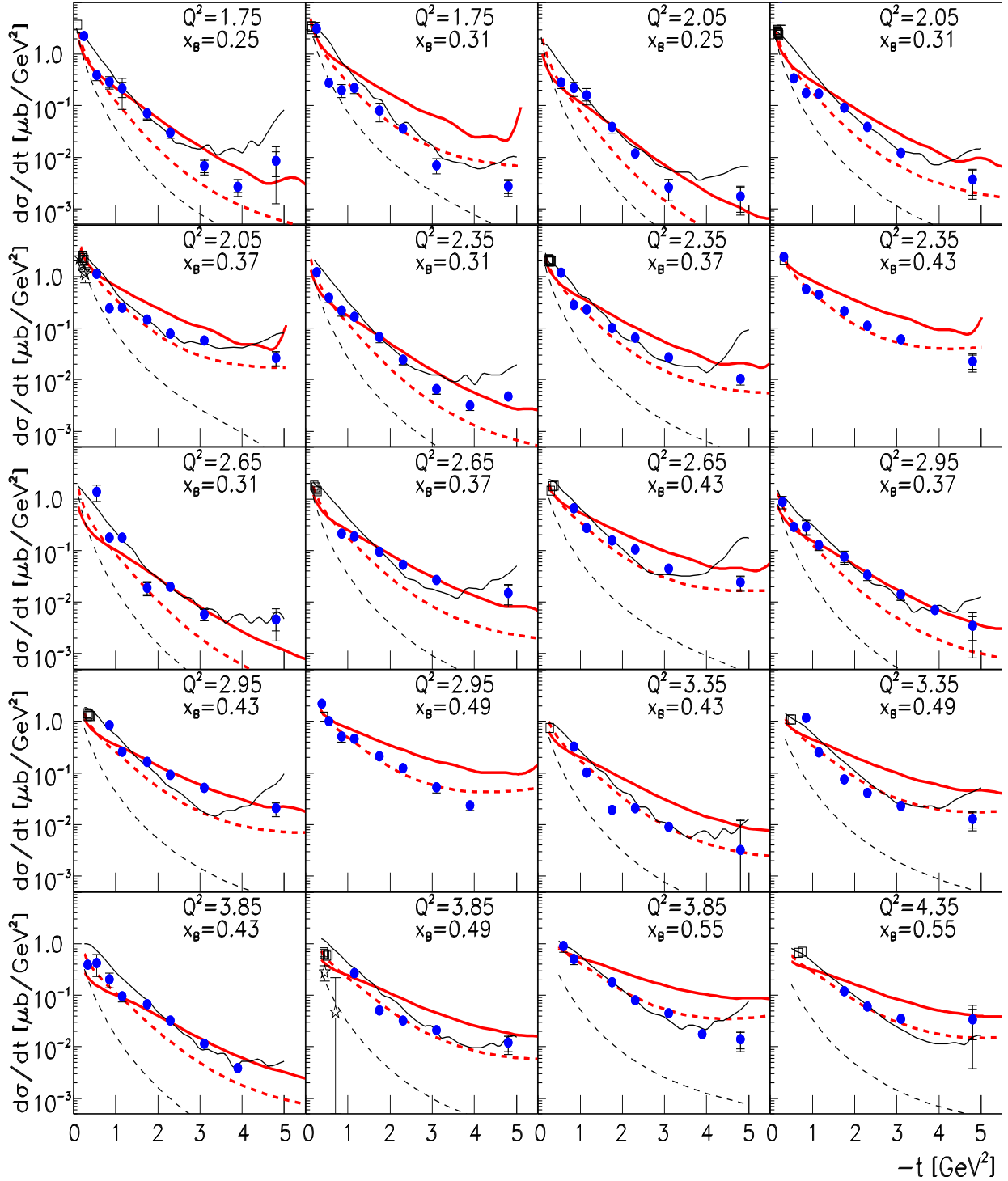}}
       \caption{
        (color online). Differential cross sections $d\sigma/dt~[\mu b/\rm{GeV^2}]$ integrated over ${\phi^*_{\pi}}$ for various ($Q^2$, $x_{B}$) bins.  The blue solid points are 
	the present work. The error bars (outer error) on all cross sections include both
statistical (inner error) and systematic uncertainties added in quadrature. 
The black open squares ($d\sigma/dt$)~\cite{XQian} and open stars ($d\sigma_L/dt$)~\cite{Horn09} are $\rm{JLab}$ $\rm{Hall~C}$ data. The red thick solid ($d\sigma/dt$), and dashed ($d\sigma_L/dt$) curves are the calculations from the Laget model~\cite{JMLaget01} with ($Q^2, t$)-dependent form factors at the photon-meson vertex. The black thin solid ($d\sigma/dt$) and dashed ($d\sigma_L/dt$) curves are the calculations from the Kaskulov {\em et al.} model~\cite{Kaskulov:2008xc}. This model does not provide the calculation at ($x_B$, $Q^2$)=(0.43, 2.35), (0.49, 2.95) due to the kinematic limit in the model.
       }
       \label{fig:crs01}
\end{figure*}

\begin{figure*}[!htb]
\vspace{210mm}
\centering{
\includegraphics{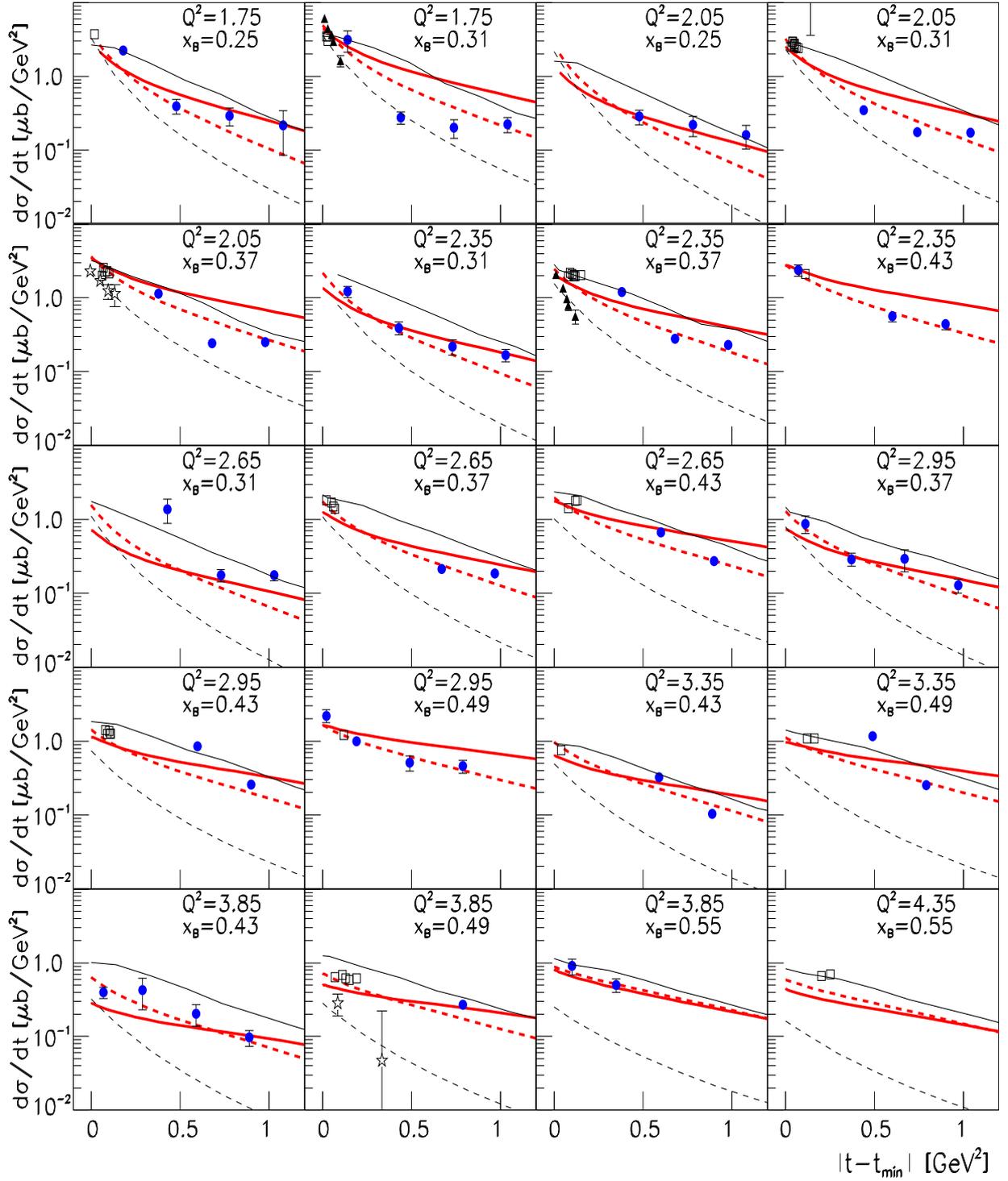}}
       \caption{
        (color online). Same as Fig.~\ref{fig:crs01} except with an expanded low $|t-t_{min}|$ scale, where -$t_{min}$ is the minimum kinematically possible four-momentum transfer.
	In addition, the black solid triangles~\cite{HPBlok} show the 
	$\rm{JLab}$ $\rm{Hall~C}$ extracted $d\sigma_L/dt$ data. 
       }
       \label{fig:crs01a}
\end{figure*}

  Fig.~\ref{fig:crs01} shows
  the differential cross section ${d\sigma}/{dt}$ as a function of $t$ 
  for different ($x_{B}$, $Q^2$) bins. We define the reduced differential cross section:
\begin{eqnarray}\label{eq:dsigmadt}
\frac{d\sigma}{dt} = \frac{1}{\Gamma}\frac{d^3\sigma}{dQ^2dx_Bdt}~, 
\end{eqnarray}
  where the virtual photon flux factor~\cite{Hand} has been factored out.
     We have included in Fig.~\ref{fig:crs01} the $\rm{JLab}$ $\rm{Hall~C}$ data, which 
     cover only the very small $t$ domain. We note that there is
  generally reasonable agreement between the results of the two experiments. However, care must be 
  taken in comparing the Hall C and Hall B measurements as the central 
     ($t$, $Q^2$, and $W$ or $x_B$, $\epsilon$) values do not exactly match each other.
For instance, the most important discrepancy
seems to appear in the bin ($x_B$, $Q^2$)=(0.49, 3.35) where the Hall C measurement 
was carried out at $\epsilon$=0.45~\cite{XQian}
while ours corresponds to $\epsilon$=0.58 (the $x_B$ and $Q^2$ values being almost similar). 
According to the value of $\sigma_L$ relative to $\sigma_T$, the Hall C cross section should then be renormalized: by a factor of 1.58/1.45$\approx$10\% (if $\sigma_L$$\approx$$\sigma_T$ which the Hall C separated data~\cite{Horn09,HPBlok} indicate, although at a slightly different kinematics) to a factor 0.58/0.45$\approx$30\% (if $\sigma_L$ dominates over $\sigma_T$ which the Laget model predicts).
For better visualization, which is also relevant for 
the comparison with
  the models, we also show Fig.~\ref{fig:crs01a} which 
  concentrates on the low $|t|$ range of Fig.~\ref{fig:crs01}.

  The ${d\sigma}/{dt}$ cross sections
  fall as $|t|$ increases, with some flattening at large
  $|t|$, which is a feature that is also observed in photoproduction~\cite{RAnderson00,LYZhu}. 
  For several bins, for instance ($x_B$, $Q^2$)=(0.31, 1.75) or (0.37, 2.05), we notice a
  structure in ${d\sigma}/{dt}$ for $|t|\approx$ 0.5 GeV$^2$. The origin of this dip
  is not known. We note that the $\rm{JLab}$ $\rm{Hall~C}$ experiment~\cite{XQian} also
  measured such a structure in ${d\sigma}/{dt}$ (see Fig.~13 in Ref.~\cite{XQian} for 
  bin ($W$, $Q^2$)=(1.8, 2.16)).

We first compare our data to calculations using hadronic degrees of 
freedom. The first one with which we will compare our data is the Laget model~\cite{JMLaget01} based on Reggeized $\pi^+$ 
and $\rho^+$ meson exchanges in the $t-$channel~\cite{VGL}. The hadronic coupling constants 
entering the calculation are all well-known or well-constrained, and the main free parameters 
are the mass scales of the electromagnetic form factors at the photon-meson vertices. 
  
  If one considers only standard, monopole, $Q^2$-dependent form 
  factors, one obtains much steeper $t$-slopes than the data.
  An agreement with the data can be recovered by introducing
  a form factor mass scale that also depends on $t$ according to the prescription
  of Ref.~\cite{JMLaget01}. This form factor accounts phenomenologically 
  for the shrinking in size of the nucleon system as $t$ increases. 
  The size of the effect is quantitatively the same as in the 
  $p(e,e^\prime \omega)p$ channel (see Fig.~1 of Ref.~\cite{JMLaget01}), which is dominated by pion exchange 
  in the same energy domain as in our study. The results of this calculation
  with ($Q^2$, $t$)-dependent meson electromagnetic form factors are shown, for 
  ${d\sigma_L}/{dt}$ and
  ${d\sigma}/{dt}={d\sigma_T}/{dt}+\epsilon{d\sigma_L}/{dt}$,  
  in Figs.~\ref{fig:crs01} and~\ref{fig:crs01a} by the red curves. The Laget model gives a qualitative
  description of the data, with respect to the overall normalization at low $t$ and the $x_B$-,
  $Q^2$- and $t$- dependencies. We recall that this model already gives
  a good description of the photoproduction data ($\rm{SLAC}$, $\rm{JLab}$) and of the 
  $\rm{HERMES}$ electroproduction data, and that the form factor mass scale~\cite{JMLaget01} 
  has not been adjusted to fit our data. 

  In the framework of this model, ${d\sigma_L}/{dt}$ dominates at low 
  $|t|$, while ${d\sigma_T}/{dt}$ takes over around 
  $|t|\approx$ 2 GeV$^2$, this value slightly varying
  from one ($Q^2$, $x_B$) bin to another. This dominance of $\sigma_L$ at low $|t|$
  is a consequence of the $t$-channel
  $\pi^+$-exchange (pion pole). At larger $|t|$, the
  $\rho^+$ meson exchange, which contributes mostly to the transverse part
  of the cross section, begins to dominate. The Laget Regge model, in addition to $t$-channel
  meson exchanges, also contains $u$-channel baryon exchanges. It thus
  exhibits an increase of the cross section in some
  ($Q^2$, $x_B$) bins at the largest $|t|$-values, corresponding to low-$|u|$ values. 
  We have additional data at larger $|t|$ (lower $|u|$) that are currently under analysis. 

The second model with which we compare our data is the ``hybrid" two-component hadron-parton model proposed in 
Refs.~\cite{Kaskulov:2008xc,Kaskulov:2009gp}. Like in the Laget model, 
it is based on the exchange of the $\pi^+$ and $\rho^+$ Regge trajectories
in the $t$-channel. However, the model complements these hadron-like 
interaction types, which dominate in photoproduction and low $Q^2$ electroproduction, 
by a direct interaction of virtual photons with partons at high values of $Q^2$
followed by string (quark) fragmentation into $\pi^+n$.  The partonic part of the production mechanism is described by a ``deep inelastic"-like 
electroproduction mechanism where the quark knockout
reaction $\gamma^* q \to q$ is followed by the 
fragmentation process of the Lund type. The transverse response is 
then treated as the exclusive limit of the semi-inclusive  reaction 
$p(e,e'\pi^+)X$. Figures~\ref{fig:crs01} and~\ref{fig:crs01a} show the results 
of this model compared to our data where very good agreement is found.
This calculation was also found to give a good description of the 
L/T-separated Hall C and unseparated HERMES data~\cite{Kaskulov:2008xc,Kaskulov:2009gp}.

  The third model that we wish to discuss, the GK model, is based purely on
   partonic degrees of freedom and is based on the handbag  GPD formalism.
   In this model ${d\sigma_L}/{dt}$ is also mostly generated by the pion pole, 
   similar to the two previous models. There are, however, a couple of differences
  in the treatment of this pion pole in the GK calculation. For instance, the Laget
  model has an intrinsic energy dependence. It is ``Reggeized", so
  the $t$-channel propagator is proportional to $s^{\alpha_\pi(t)}$, 
  where $\alpha_{\pi}(t)$ is the pion Regge trajectory. In addition, it uses 
  a ($Q^2$, $t$)-dependent 
  electromagnetic form factor. These two features change the
  $s$-, $x_B$-, and $t$- dependencies of the pion pole with respect to
  the GK treatment. Indeed, in the latter case, the $t$-channel pion propagator is
  proportional to ${1}/{(t-m^2_\pi)}$, so it has no s-dependence, and 
  the hadronic form factor at the $\pi NN$ vertex is only $t$-dependent.

\begin{figure*}[!htb]
\vspace{210mm}
\centering{
\includegraphics{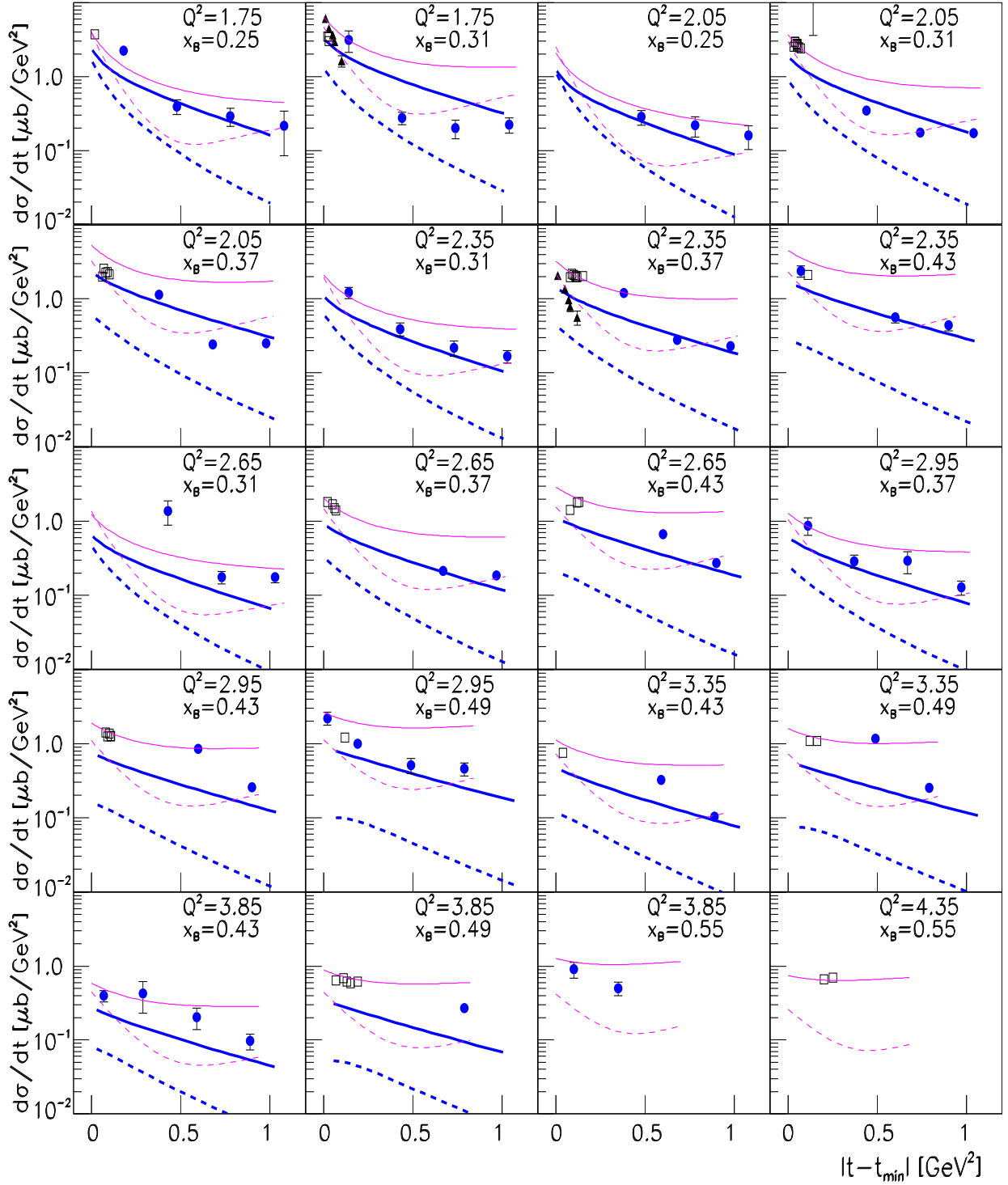}}
       \caption{
        (color online). Same as Fig.~\ref{fig:crs01a} except with $|t-t_{min}|$ scale. The blue thick solid and dashed curves are $d\sigma$/$dt$ and $d\sigma_L$/$dt$, respectively, from the GK model~\cite{GK11}. The magenta thin solid ($d\sigma/dt$) and dashed ($d\sigma_L/dt$) curves are the calculations from the 2nd Kaskulov {\em et al.} model~\cite{Kaskulov:2010kf}.
       }
       \label{fig:crs01a_bis}
\end{figure*}

 Figure~\ref{fig:crs01a_bis} shows the results 
  of the GK calculation (in blue) for ${d\sigma_L}/{dt}$ and ${d\sigma}/{dt}$.
  We recall that the GK model is applicable 
  only for small values of $-t/Q^2$. Outside this regime, higher-twist 
  contributions that are not taken into account in the GK handbag formalism 
  are expected.  
  The GK model describes qualitatively our low-$t$ unseparated cross sections over
  our whole ($x_B$, $Q^2$) domain. This is remarkable since
  the GK model was optimized for higher-energy kinematics ($\rm{HERMES}$)
  and no further adjustments were made for the present $\rm{CLAS}$ kinematics.
   We see that ${d\sigma_L}/{dt}$ has a non-negligible contribution only in the
  very low $|t|$ domain and only for a few ($x_B$, $Q^2$) bins, in particular at the
  lowest $x_B$ and the largest $Q^2$ values (for instance, the ($x_B$, $Q^2$) bins (0.25, 1.75)
  and (0.31, 2.35)). This is in line with the
  observation that at $\rm{HERMES}$ kinematics, i.e. at lower $x_B$ and larger $Q^2$ values,
  the longitudinal part of the cross section dominates in the GK model at low $|t|$.
  For the larger $x_B$ values,  one sees that the 
    dominance of ${d\sigma_L}/{dt}$ at low $|t|$ is not at all
  systematic in the GK calculation. The ratio of ${d\sigma_L}/{dt}$ to 
  ${d\sigma}/{dt}$ strongly
  depends on $x_B$. Specifically, it decreases as $x_B$ increases and
  at $x_B$=0.49, ${d\sigma_L}/{dt}$ is only a few percent of ${d\sigma}/{dt}$,
  even at the lowest $t$ values. This is a notable difference from the 
  Laget Regge model for instance.
%/data/scratch/parkkj/ana_note_scaling/review_plot_fig39/crsplotmodel2012combinefinalwithhallc4e.kumac

% exe : CrossSectionFinalwithAcc2012_NOV.tmin.exe
%/media/_local/scratch/parkkj/ana_note_scaling/review_plot_fig39/plot_epja_fig14e.kumac

% exe : CrossSectionFinalwithAcc2012_NOV.tmin.exe
% crsplotmodel2012combinefinalwithhallczoom4ctmin2e.kumac
\clearpage
 
In particular, one can remark in Fig.~\ref{fig:crs01a_bis}, where we display
  in two ($x_B$, $Q^2$) bins ((0.31, 1.75) and (0.37, 2.35)) the 
  longitudinal part of the cross section as extracted from Hall C~\cite{HPBlok},
  that the longitudinal part of the GK calculation is not in good agreement 
  with the experimental data. This can be attributed to the way the pion pole
  and/or the pion-nucleon form factor, which are the main contributors
  to the longitudinal part of the cross section, are modeled in the GK approach.
  A Reggeization (like in the Laget model) or a change in the pion-nucleon form factor 
  parametrization could possibly enhance the pion pole contribution at JLab kinematics
  and provide better agreement with our data (without damaging the agreement with the 
  HERMES data)~\cite{private_kroll}. We recall that the GK model for which the GPD
  parameters were fitted to the low $x_B$ HERMES data, was simply extrapolated to 
  the kinematics of the present article without any optimization and thus the present 
  disagreement observed in ${d\sigma_L}/{dt}$ should not be considered as definitive.

  In the GK model, the transverse part of the cross
  section is due to transversity GPDs. In Fig.~\ref{fig:crs01a_bis}, the 
  GK calculation predicts that the transverse 
  part of the cross section dominates essentially everywhere in our kinematic 
  domain. Although the GK L/T ratio probably needs to be adjusted as
  we just discussed, the GK calculation opens the original and exciting 
  perspective to access transversity GPDs through exclusive $\pi^+$
  electroproduction.

Finally, at the kinematics of our experiment, in spite of our
$W>2$ GeV cut, it cannot be excluded that nucleon resonances
contribute. In Ref.~\cite{Kaskulov:2010kf},
Kaskulov and Mosel identify these high-lying resonances with partonic excitations 
in the spirit of the
resonance-parton duality hypothesis and invoke the continuity in going from 
an inclusive final deep inelastic state to exclusive pion production. During this 
transition one expects that the inclusion of resonance excitations enhances the 
transverse response while leaving the longitudinal strength originating in the 
$t$-channel meson exchanges intact. Thus, in this work, the $t$-channel exchange 
part of the production amplitude is again described by the exchange of the
Regge trajectories ($\pi^+$, $\rho^+$ and $a_1^+$) to which it is added
a nucleon resonance component that is described via
a dual connection between the resonance and partonic deep inelastic processes. 
The parameters of this model have been tuned using the forward JLab Hall C data. 
  Figure~\ref{fig:crs01a_bis} shows the results 
 of this calculation with our data and a reasonable agreement is found.
 
 The four models that we just discussed, although they give a reasonable
 description of the unseparated cross sections, display rather different L/T ratios.
 The precise measurement of this ratio as a function of $x_B$, $Q^2$ and $t$
 appears thus as essential to clarify the situation.
  For instance,  in order to validate and/or tune the GK approach, it would be interesting 
  to study the $Q^2$-dependence at fixed $x_B$ and $t$ 
  of the longitudinal and transverse cross sections. They should approach, as 
  $Q^2$ increases, a ${1}/{Q^6}$ and ${1}/{Q^8}$ scaling behavior respectively,
  as mentioned in the introduction of this article.
  In contrast, the Laget Regge model, for which $x_B$ is not a ``natural" variable
  (it is rather $W$) should not predict such a $Q^2$ scaling at fixed $x_B$. 
  Although we are probably very far from such an asymptotic regime,
  the measurement of the $Q^2$-dependence in the transition region accessible
  with the upcoming $\rm{JLab}$ 12-GeV upgrade should provide some
  strong constraints and in particular some checks on the way the higher-twist 
  corrections are treated in the GK model. Such a program is already planned
  at JLab~\cite{hallc-12gevprop}. %"

\subsection{${d\sigma}/dt$ as a function of $Q^2$ at fixed $t$}
\begin{figure*}[htb]
\begin{center}
        \includegraphics[angle=0,width=0.7\textwidth]{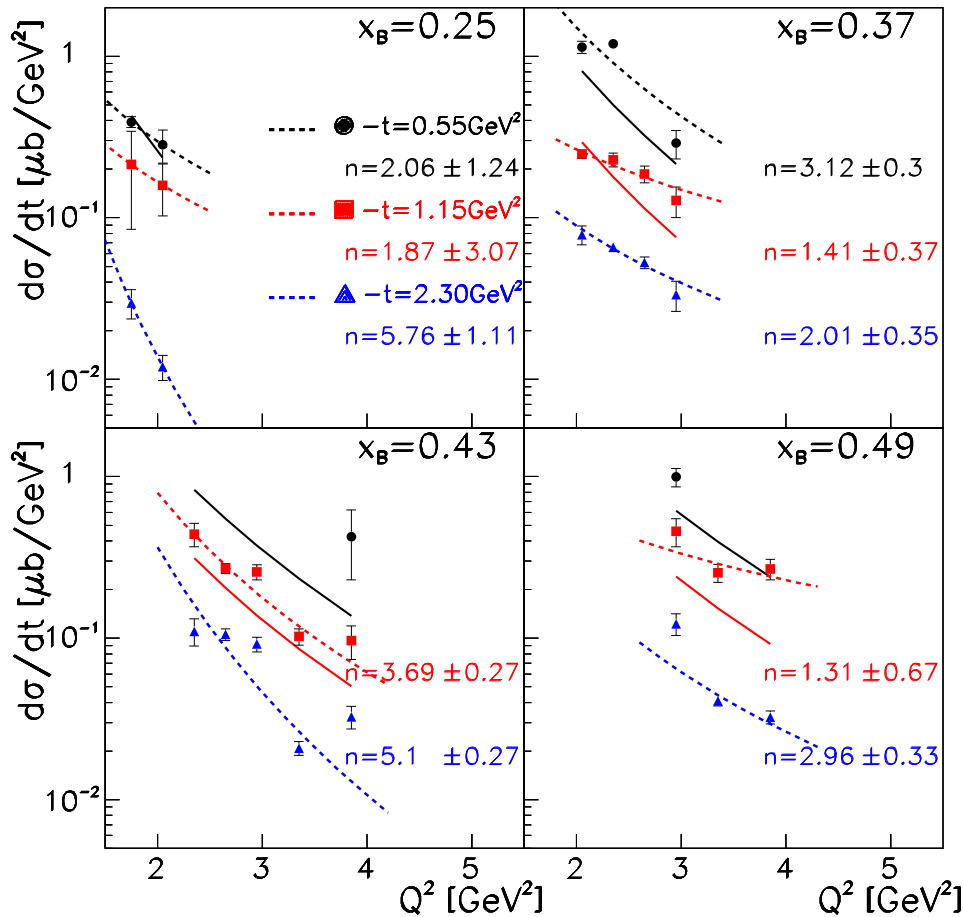}
        \caption{
          (color online). Differential cross sections $d\sigma/dt\;[\rm{\mu b/GeV^2}]$ 
	  versus $Q^2$ at fixed $x_{B}$ for various $t$ values. The
	  dashed curves are the results of a fit to the function $A/Q^n$. 
	  The solid curves are the results of the GK calculations~\cite{GK11}.
	  The GK calculations are only valid for $-t <\; \approx1$ GeV$^2$ so
	  we do not display those results
	  for $-t = 2.3$ GeV$^2$. 
	             \label{fig:q2_scaling_fig1}
        }
\end{center}
\end{figure*}

\begin{figure*}[htb]
\begin{center}
        \includegraphics[angle=0,width=0.7\textwidth]{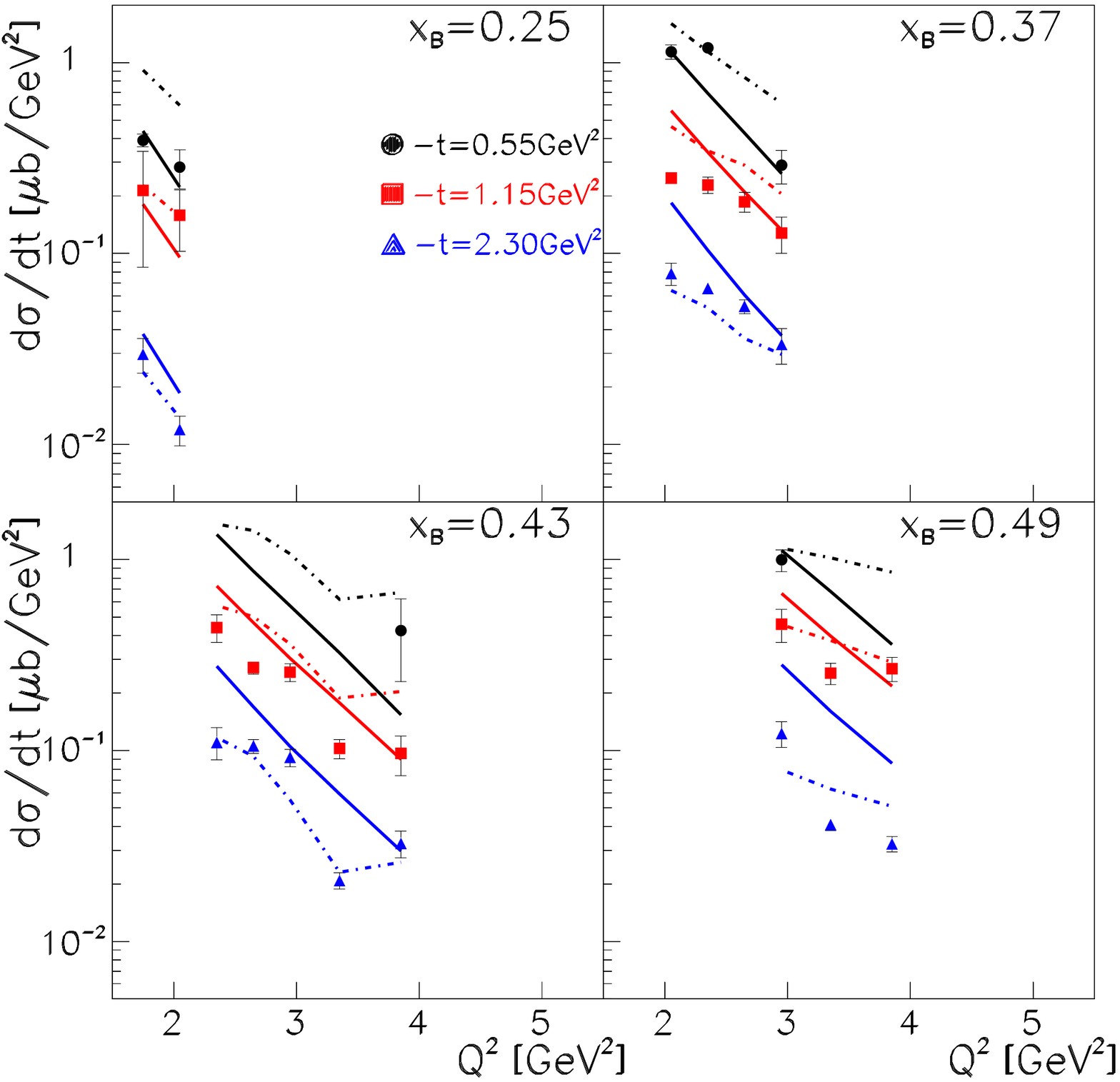}
        \caption{
          (color online). Differential cross sections $d\sigma/dt\;[\rm{\mu b/GeV^2}]$ 
	  versus $Q^2$ at fixed $x_{B}$ for various $t$ values. 
	  The solid curves are the results of the Laget calculations~\cite{JMLaget01}
	  and the dash-dotted curves of the ``hybrid" two-component hadron-parton model 
	  of Refs.~\cite{Kaskulov:2008xc,Kaskulov:2009gp}.
	             \label{fig:q2_scaling_fig2}
        }
\end{center}
\end{figure*}

Figures~\ref{fig:q2_scaling_fig1} and ~\ref{fig:q2_scaling_fig2} show the differential cross section $d\sigma$/$dt$
as a function of $Q^2$ at fixed $x_B$ for various $t$ values. 
In Fig.~\ref{fig:q2_scaling_fig1}, our data
are fitted with a $1/Q^n$ function and are compared to the GK model. We recall that,
at asymptotically large $Q^2$, the handbag mechanism predicts a dominance
of $\sigma_L$ which should scale as ${1}/{Q^6}$ at fixed $t$ and $x_B$. The resulting exponents $n$ 
of our fit indicates a flatter $Q^2$ dependence than $1/Q^6$. At the relatively
low $Q^2$ range accessed in this experiment, higher-twist effects are expected to contribute and 
hence the leading-twist ${1}/{Q^6}$ dependence of $\sigma_L$ is no longer expected.
We note that such higher-twist contributions are part of the GK calculation
and the GK model also does not show this scaling behavior at the present $Q^2$ values.
Although the GK model tends to underestimate the normalization
of our data, its $Q^2$ dependence agrees reasonably well with our data.

In Fig.~\ref{fig:q2_scaling_fig2}, we compare our data to the Laget~\cite{JMLaget01}
and the Kaskulov {\em et al.}~\cite{Kaskulov:2008xc,Kaskulov:2009gp} models.
The Laget calculation gives a reasonable description of the 
data although it seems to have a slightly steeper $Q^2$-dependence 
than our data (particularly in the $x_B$=0.37 bin). 
We note that in the $x_B$=0.43 bin, our data seem to display a structure (dip)
for $Q^2$ values between 3 and 4 GeV$^2$, which is certainly intriguing.
We have at this stage no particular explanation for this. We just observe
that the ``hybrid" two-component hadron-parton model of Refs.~\cite{Kaskulov:2008xc,Kaskulov:2009gp} displays apparently also
such structure which should therefore be further investigated.

% dir : /media/_local/scratch/parkkj/ana_note_scaling/review_plot_fig39/q2scaling3.prc.dsdt.opt3ee.kumac

\subsection{${d\sigma}/{dt}$ as a function of $W$ at fixed $\theta_{\pi}^*$}

\begin{figure*}[htb]
\begin{center}
        \includegraphics[angle=0,width=0.7\textwidth]{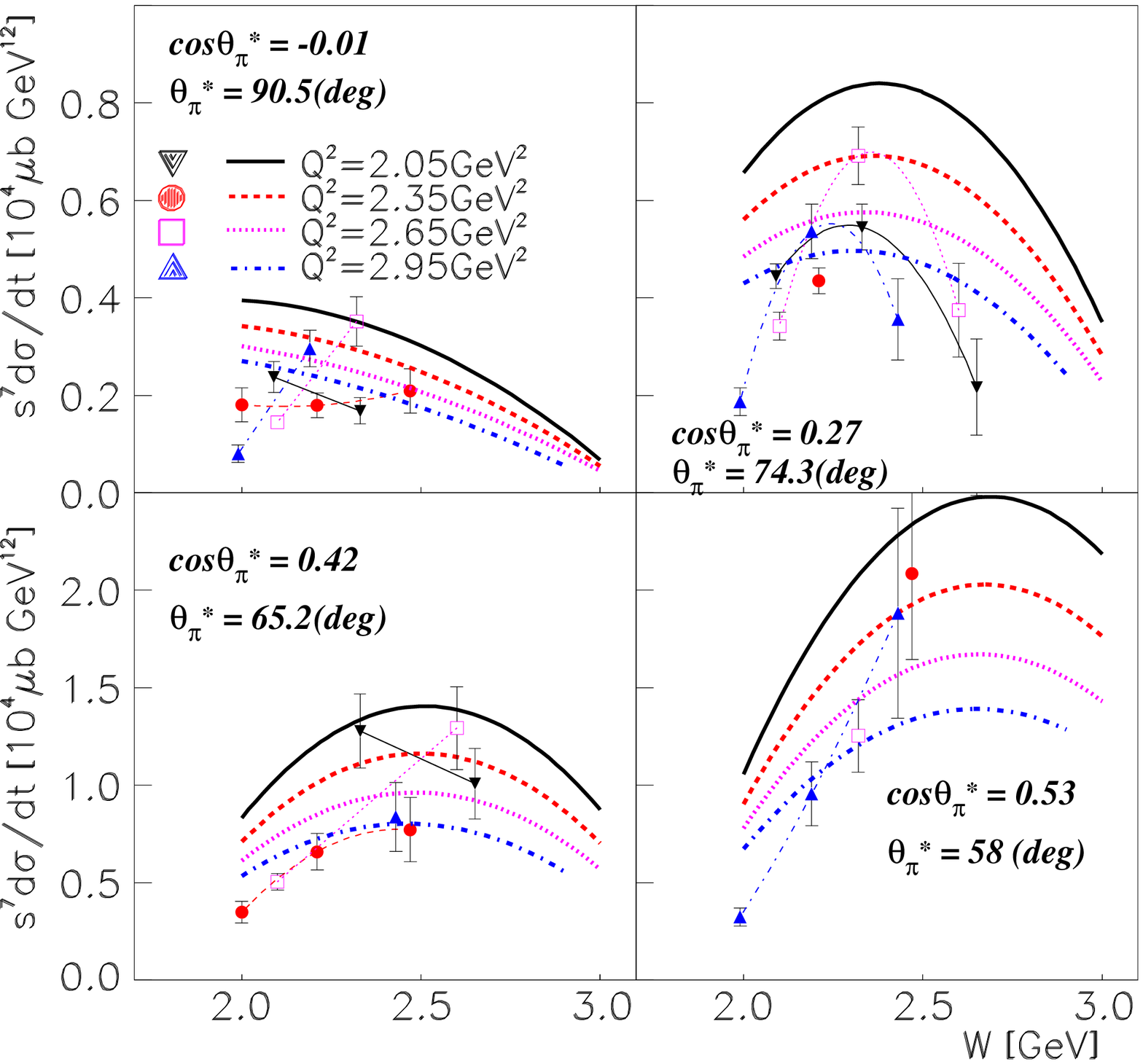}
        \caption{
          (color online). Scaled cross sections $s^{7}d\sigma/dt$ $[10^4\rm{\mu b}\;\rm{GeV^{12}}]$ versus $W$ 
	  for $\theta_{\pi}^* \approx$ 60$^\circ \ldots$ 90$^\circ$ and various $Q^2$ bins. 
	  Thick curves are from the Laget model~\cite{JMLaget01}. The thin curves are to guide the eye, connecting points with the same $Q^2$ values.
	             \label{fig:scaled_crs0}
        }
\end{center}
\end{figure*}

 Figure~\ref{fig:scaled_crs0} shows our scaled cross sections, $s^7 d\sigma/dt$, as a 
 function of  $W$ for four $Q^2$ values and 
 four bins in $\cos\theta_\pi^*$: $-0.01\pm 0.16$, $0.27\pm0.1$, $0.42\pm0.05$ and
 $0.53\pm0.06$. The lever arm in $W$ is limited. At $\theta_{\pi}^*=90^{\circ}$, 
 where the scaling behavior is expected to set in most quickly, we have 
 only 2 or 3 data points in $W$, depending on the $Q^2$ bin.
 It is therefore difficult to draw precise conclusions at this stage 
 for the $W$-dependence at fixed $Q^2$.
 Nevertheless, with these limited (but unique)
 data, one can say that, at $\theta_{\pi}^*=90^\circ$, except for the 3 data points at $Q^2$=2.35 GeV$^2$, 
 the $W$-dependence of $s^7 {d\sigma}/{dt}$ does not appear to be constant. 
 We also display in Fig.~\ref{fig:scaled_crs0} the result of the Laget model. 
   It gives, within a factor two, a general
description of these large-angle data. The
$W$-dependence of our data is very similar to the energy dependence that was observed in
photoproduction~\cite{WChen}. In the same energy range as covered by 
the present study, real-photon data exhibit strong deviations from 
scaling. Within the Laget model,
these deviations are accounted 
for by the coupling between the $n \pi^+$ and the $\rho N$ 
channels~\cite{JML2010}. The $\rm{JLab}$ 12-GeV upgrade will allow us to increase the
 coverage in $W$ and check whether the hints of oscillations that we observe remain in the 
virtual-photon sector.

%/media/_local/scratch/parkkj/ana_note_scaling/scaling_review2011/sscalingexamplot2012prcmodel2.kumac

\section{Summary}
 We have measured the cross sections of 
 exclusive electroproduction of $\pi^+$ mesons from protons as a function 
 of $-t=0.1$ - $5.3\;\rm{GeV^2}$, $x_B=0.16$ - $0.58$, and $Q^2=1.6$ - $4.5\;\rm{GeV^2}$. 
 We have compared our differential cross sections to four recent
 calculations based on hadronic and
 partonic degrees of freedom. The four models give a qualitative
 description of the overall strength and of the $t$-, $Q^2$- and $x_B$-
 dependencies of our unseparated cross sections. There is an obvious need
 for L-T separated cross sections in order to distinguish
 between the several approaches. These separations will be 
 possible with the upcoming JLab 12-GeV upgrade. In particular, if the handbag approach can accomodate
 the data, the $p(e, e^{\prime}\pi^+)n$ process offers the outstanding potential
 to access transversity GPDs.

\vspace{5mm} 

\section{Acknowledgment}
 We acknowledge the outstanding efforts of the staff of the Accelerator and the Physics Divisions at Jefferson Lab that made this experiment possible. We also give many thanks to P.~Kroll, S.~Goloskokov and M.~Kaskulov for their calculations. The early work of D.~Dor\'e on this analysis is also acknowledged. This work was supported in part by the US Department of Energy, the National Science Foundation, the Italian Istituto Nazionale di Fisica Nucleare, the French American Cultural Exchange (FACE) and Partner University Funds (PUF) programs,
the French Centre National de la Recherche Scientifique, the French Commissariat \`a l'Energie Atomique, the United Kingdom's Science and Technology Facilities Council, the Chilean Comisi\'on Nacional de Investigaci\'on Cient\'ifica y Tecnol\'ogica (CONICYT), and the National Research Foundation of Korea. The Southeastern Universities Research Association (SURA) operated the Thomas Jefferson National Accelerator Facility for the US Department of Energy under Contract No.DE-AC05-84ER40150.

%
%
%
%

%%%%%%%%%%%%%%%%%%  References    %%%%%%%%%%%%%%%%%%%%%%%%%%%

\end{document}

%% file: authors06112012.tex
\newcommand*{\ANL}{Argonne National Laboratory, Argonne, Illinois 60439, USA}
\newcommand*{\ANLindex}{1}

\newcommand*{\ASU}{Arizona State University, Tempe, Arizona 85287-1504, USA}
\newcommand*{\ASUindex}{2}

\newcommand*{\CSUDH}{California State University, Dominguez Hills, Carson, California 90747, USA}
\newcommand*{\CSUDHindex}{3}

\newcommand*{\CANISIUS}{Canisius College, Buffalo, New York 14208, USA}
\newcommand*{\CANISIUSindex}{4}

\newcommand*{\CMU}{Carnegie Mellon University, Pittsburgh, Pennsylvania 15213, USA}
\newcommand*{\CMUindex}{5}

\newcommand*{\CUA}{Catholic University of America, Washington, D.C. 20064, USA}
\newcommand*{\CUAindex}{6}

\newcommand*{\SACLAY}{CEA, Centre de Saclay, Irfu/Service de Physique Nucl\'eaire, 91191 Gif-sur-Yvette, France}
\newcommand*{\SACLAYindex}{7}

\newcommand*{\CNU}{Christopher Newport University, Newport News, Virginia 23606, USA}
\newcommand*{\CNUindex}{8}

\newcommand*{\UCONN}{University of Connecticut, Storrs, Connecticut 06269, USA}
\newcommand*{\UCONNindex}{9}

\newcommand*{\EDINBURGH}{Edinburgh University, Edinburgh EH9 3JZ, United Kingdom}
\newcommand*{\EDINBURGHindex}{10}

\newcommand*{\FU}{Fairfield University, Fairfield Connecticut 06824, USA}
\newcommand*{\FUindex}{11}

\newcommand*{\FIU}{Florida International University, Miami, Florida 33199, USA}
\newcommand*{\FIUindex}{12}

\newcommand*{\FSU}{Florida State University, Tallahassee, Florida 32306, USA}
\newcommand*{\FSUindex}{13}

\newcommand*{\Genova}{Universit$\grave{a}$ di Genova, 16146 Genova, Italy}
\newcommand*{\Genovaindex}{14}

\newcommand*{\GWUI}{The George Washington University, Washington, D.C. 20052, USA}
\newcommand*{\GWUIindex}{15}

\newcommand*{\ISU}{Idaho State University, Pocatello, Idaho 83209, USA}
\newcommand*{\ISUindex}{16}

\newcommand*{\INFNFE}{INFN, Sezione di Ferrara, 44100 Ferrara, Italy}
\newcommand*{\INFNFEindex}{17}

\newcommand*{\INFNFR}{INFN, Laboratori Nazionali di Frascati, 00044 Frascati, Italy}
\newcommand*{\INFNFRindex}{18}

\newcommand*{\INFNGE}{INFN, Sezione di Genova, 16146 Genova, Italy}
\newcommand*{\INFNGEindex}{19}

\newcommand*{\INFNRO}{INFN, Sezione di Roma Tor Vergata, 00133 Rome, Italy}
\newcommand*{\INFNROindex}{20}

\newcommand*{\ORSAY}{Institut de Physique Nucl\'eaire ORSAY, Orsay, France}
\newcommand*{\ORSAYindex}{21}

\newcommand*{\ITEP}{Institute of Theoretical and Experimental Physics, Moscow, 117259, Russia}
\newcommand*{\ITEPindex}{22}

\newcommand*{\JMU}{James Madison University, Harrisonburg, Virginia 22807, USA}
\newcommand*{\JMUindex}{23}

\newcommand*{\KNU}{Kyungpook National University, Daegu 702-701, Republic of Korea}
\newcommand*{\KNUindex}{24}

\newcommand*{\LPSC}{LPSC, Universite Joseph Fourier, CNRS/IN2P3, INPG, Grenoble, France}
\newcommand*{\LPSCindex}{25}

\newcommand*{\UNH}{University of New Hampshire, Durham, New Hampshire 03824-3568, USA}
\newcommand*{\UNHindex}{26}

\newcommand*{\NSU}{Norfolk State University, Norfolk, Virginia 23504, USA}
\newcommand*{\NSUindex}{27}

\newcommand*{\OHIOU}{Ohio University, Athens, Ohio  45701, USA}
\newcommand*{\OHIOUindex}{28}

\newcommand*{\ODU}{Old Dominion University, Norfolk, Virginia 23529, USA}
\newcommand*{\ODUindex}{29}

\newcommand*{\RPI}{Rensselaer Polytechnic Institute, Troy, New York 12180-3590, USA}
\newcommand*{\RPIindex}{30}

\newcommand*{\URICH}{University of Richmond, Richmond, Virginia 23173, USA}
\newcommand*{\URICHindex}{31}

\newcommand*{\ROMAII}{Universita' di Roma Tor Vergata, 00133 Rome, Italy}
\newcommand*{\ROMAIIindex}{32}

\newcommand*{\MSU}{Skobeltsyn Nuclear Physics Institute, Skobeltsyn Nuclear Physics Institute, 119899 Moscow, Russia}
\newcommand*{\MSUindex}{33}

\newcommand*{\SCAROLINA}{University of South Carolina, Columbia, South Carolina 29208, USA}
\newcommand*{\SCAROLINAindex}{34}

\newcommand*{\JLAB}{Thomas Jefferson National Accelerator Facility, Newport News, Virginia 23606, USA}
\newcommand*{\JLABindex}{35}

\newcommand*{\UTFSM}{Universidad T\'{e}cnica Federico Santa Mar\'{i}a, Casilla 110-V Valpara\'{i}so, Chile}
\newcommand*{\UTFSMindex}{36}

\newcommand*{\GLASGOW}{University of Glasgow, Glasgow G12 8QQ, United Kingdom}
\newcommand*{\GLASGOWindex}{37}

\newcommand*{\VT}{Virginia Polytechnic Institute and State University, Blacksburg, Virginia   24061-0435, USA}
\newcommand*{\VTindex}{38}

\newcommand*{\VIRGINIA}{University of Virginia, Charlottesville, Virginia 22901, USA}
\newcommand*{\VIRGINIAindex}{39}

\newcommand*{\WM}{College of William and Mary, Williamsburg, Virginia 23187-8795, USA}
\newcommand*{\WMindex}{40}

\newcommand*{\YEREVAN}{Yerevan Physics Institute, 375036 Yerevan, Armenia}
\newcommand*{\YEREVANindex}{41}

\newcommand*{\NOWMSU}{Skobeltsyn Nuclear Physics Institute, 119899 Moscow, Russia}
\newcommand*{\NOWORSAY}{Institut de Physique Nucl\'eaire ORSAY, Orsay, France}
\newcommand*{\NOWINFNGE}{INFN, Sezione di Genova, 16146 Genova, Italy}
\newcommand*{\NOWROMAII}{Universita' di Roma Tor Vergata, 00133 Rome, Italy}
 %%%%%%%%%%%%%%% END OF Latex Macros for institute addresses  %%%%%%%%%%%%%%%%%%%%%%%%% 20 opt-out
\author{ K.~Park   \inst{\JLABindex}  
\and  M.~Guidal   \inst{\ORSAYindex}  
\and  R.W.~Gothe   \inst{\SCAROLINAindex}  
\and  J.M.~Laget   \inst{\JLABindex}
\and  M.~Gar\c con \inst{\SACLAYindex}
\and  K.P. ~Adhikari   \inst{\ODUindex}  
\and  M.~Aghasyan   \inst{\INFNFRindex}  
\and  M.J.~Amaryan  \inst{\ODUindex}  
\and  M.~Anghinolfi   \inst{\INFNGEindex} 
\and  H.~Avakian   \inst{\JLABindex}
% ...
}

%\and  \Red{D.~Adijaram}  \inst{\ODUindex}  

%\and  \Red{V.~Batourine}  \inst{\JLABindex}  

%\and  \Red{D.~Branford}  \inst{\EDINBURGHindex}  

%\and \Red{P.L.~Cole}  \inst{\ISUindex}   \fnmsep  \inst{\JLABindex}  
%\and  \Red{V.~Crede}   \inst{\FSUindex}  
%\and  \Red{A.~Fradi}  \inst{\ORSAYindex}  
%% \and  M.Y.~Gabrielyan \inst{\FIUindex}
%\and  \Red{J.T.~Goetz} \inst{\UCLAindex}
%\and  \Red{Y.~Gotra}   \inst{\JLABindex}  

%\and  \Red{C.E.~Hyde} \inst{\ODUindex}
%\and  \Red{R.~Nasseripour} \inst{\GWUIindex} \fnmsep  \inst{\FIUindex}
%\and  \Red{Y.~Prok} \inst{\CNUindex} \fnmsep  \inst{\VIRGINIAindex}
%\and  \Red{B.G.~Ritchie} \inst{\ASUindex}
%\and  \Red{A.V.~Vlassov} \inst{\ITEPindex}
%\and  \Red{V.~Ziegler}   \inst{\JLABindex}  

\institute{
\ANL
\and
\ASU
\and
\CSUDH
\and
\CANISIUS
\and
\CMU
\and
\CUA
\and
\SACLAY
\and
\CNU
\and
\UCONN
\and
\EDINBURGH
\and
\FU
\and
\FIU
\and
\FSU
\and
\Genova
\and
\GWUI
\and
\ISU
\and
\INFNFE
\and
\INFNFR
\and
\INFNGE
\and
\INFNRO
\and
\ORSAY
\and
\ITEP
\and
\JMU
\and
\KNU
\and
\LPSC
\and
\UNH
\and
\NSU
\and
\OHIOU
\and
\ODU
\and
\RPI
\and
\URICH
\and
\ROMAII
\and
\MSU
\and
\SCAROLINA
\and
\JLAB
\and
\UTFSM
\and
\GLASGOW
\and
\VT
\and
\VIRGINIA
\and
\WM
\and
\YEREVAN
} 

%% file: KPark_EPJA_DVMP_Piplus.accepted.bbl
\begin{thebibliography}{0}
\bibitem{RAnderson00} R. L. Anderson {\em et al.}, Phys. Rev. D \textbf{14}, 679 (1976); C. White {\em et al.}, Phys. Rev. D \textbf{49}, 58 (1994).
\bibitem{WChen} W. Chen {\em et al.}, Phys. Rev. Lett. \textbf{103}, 012301 (2009).
\bibitem{bebek76} C. J. Bebek {\em et al.}, Phys. Rev. D {\bf 13}, 25  (1976).
\bibitem{bebek78} C. J. Bebek {\em et al.}, Phys. Rev. D {\bf 13}, 1693  (1978).
\bibitem{Horn09} T. Horn {\em et al.}, Phys. Rev. C \textbf{78}, 058201 (2008).
%% -5
\bibitem{HPBlok} H. P. Blok {\em et al.}, Phys. Rev. C \textbf{78}, 045202 (2008).
\bibitem{XQian} X. Qian {\em et al.}, Phys. Rev. C \textbf{81}, 055209 (2010).
%% - 15
\bibitem{Hermes} A. Airapetian {\em et al.}, Phys. Lett. B \textbf{659}, 486 (2008).
\bibitem{muller} D. M\"uller, D. Robaschik, B. Geyer, F.-M. Dittes, and J. Ho$\check{r}$ej$\check{s}$i, Fortschr. Phys. {\bf 42}, 101 (1994).
\bibitem{ji} X. Ji, Phys. Rev. Lett. {\bf 78}, 610 (1997); Phys. Rev. D {\bf 55}, 7114 (1997).
\bibitem{rady} A.V. Radyushkin, Phys. Lett. B {\bf 380} (1996) 417; Phys. Rev. D {\bf 56}, 5524 (1997).
\bibitem{JCCollins} J. C. Collins, L. Frankfurt, and M. Strikman, Phys. Rev. D \textbf{56}, 2982 (1997).
%% -10
\bibitem{GK09} S. V. Goloskokov and P. Kroll, Eur. Phys. J. C \textbf{65}, 137 (2010).
\bibitem{lepage} S. J. Brodsky and G. P. Lepage, Phys. Rev. D \textbf{22}, 2157 (1980).
\bibitem{SBrodsky00} S. J. Brodsky and G. R. Farrar, Phys. Rev. Lett. \textbf{31}, 1153 (1973); Phys. Rev. D \textbf{11}, 1309 (1975); V. Matveev {\em et al.}, Nuovo Cimento Lett. \textbf{7}, 719 (1973).
\bibitem{VGG00} M. Vanderhaeghen, P. A. M Guichon, and M. Guidal, Phys. Rev. D \textbf{60}, 094017 (1999).
\bibitem{GK11} S. V. Goloskokov and P. Kroll, Eur. Phys. J. A \textbf{47}, 112 (2011).
\bibitem{manki} L. Mankiewicz, G. Piller and A. Radyushkin, Eur. Phys. J. C \textbf{10}, 307 (1999).
\bibitem{franki} L. Frankfurt, P. Pobylitsa, M. Poliakov, and M. Strikman, Phys. Rev. D \textbf{60}, 014010 (1999).
\bibitem{LYZhu} L. Y. Zhu {\em et al.}, Phys. Rev. Lett. \textbf{91}, 022003 (2003), Phys. Rev. C \textbf{71}, 044603 (2005).
%% - 20
\bibitem{JNapolitano00} J. Napolitano {\em et al.}, Phys. Rev. Lett. \textbf{61}, 2530 (1988); S. J. Freedman {\em et al.}, Phys. Rev. C \textbf{48}, 1864 (1993); J. E. Belz {\em et al.}, Phys. Rev. Lett. \textbf{74}, 646 (1995).
\bibitem{CBochna00} C. Bochna {\em et al.}, Phys. Rev. Lett. \textbf{81}, 4576 (1998).
\bibitem{ESchulte00} E. C. Schulte {\em et al.}, Phys. Rev. Lett. \textbf{87}, 102302 (2001).
\bibitem{PRossi00} P. Rossi {\em et al.}, Phys. Rev. Lett. \textbf{94}, 012301 (2005); M. Mirazita {\em et al.}, Phys. Rev. C \textbf{70}, 014005 (2004).
\bibitem{Schumacher:2010qx} R.~A.~Schumacher and M.~M.~Sargsian,
  %``Scaling and Resonances in Elementary K^+ Lambda Photoproduction,''
  Phys.\ Rev.\ C {\bf 83}, 025207 (2011).
  %%[arXiv:1012.2126 [hep-ph]].
  %%CITATION = ARXIV:1012.2126;%%
%% - 25
\bibitem{CLAS} B.A. Mecking {\em et al.}, Nucl. Instrum. Methods A \textbf{51}, 409 (1995).
\bibitem{KPark00} K. Park {\em et al.}, Phys. Rev. C \textbf{77}, 015208 (2008). 
\bibitem{genev} P. Corvisiero {\em et al.}, Nucl. Instrum. Methods A \textbf{346} 433 (1994) and 
E. Golovach, M. Ripani, M. Battaglieri, R. De Vita, private communication.
\bibitem{MoTsai00} L. W. Mo, Y. S. Tsai, Rev. Mod. Phys. \textbf{41}, 205 (1969).
\bibitem{fsgen} S. Stepanyan, private communication.
%% - 30
\bibitem{afanasev} A. Afanasev {\em et al.},  Phys. Rev. D \textbf{66}, 074004 (2002).
\bibitem{Hand} L. Hand, Phys. Rev. \textbf{129}, 1834 (1963).
\bibitem{JMLaget01} J. M. Laget, Phys. Rev. D \textbf{70}, 054023 (2004). 
\bibitem{VGL} M. Guidal, J. M. Laget and M. Vanderhaeghen, Nucl. Phys. A \textbf{627}, 645 (1997), Phys. Lett. B \textbf{400}, 6 (1997). 
\bibitem{Kaskulov:2008xc} M.~M.~Kaskulov, K.~Gallmeister and U.~Mosel, Phys.\ Rev.\  D {\bf 78}, 114022 (2008). % fragmentation model
\bibitem{Kaskulov:2009gp} M.~M.~Kaskulov and U.~Mosel, Phys.\ Rev.\ C {\bf 80}, 028202 (2009). % fragmentation model ...some prediction for 12GeV....
\bibitem{private_kroll} P. Kroll, private communication.
\bibitem{Kaskulov:2010kf} M.~M.~Kaskulov and U.~Mosel, Phys.\ Rev.\ C {\bf 81}, 045202 (2010). % resonance model
\bibitem{hallc-12gevprop} T. Horn {\em et al.}, Jefferson Lab E12-07-105, \url{http://www.jlab.org/exp_prog/proposals/07/PR12-07-105.pdf}.
\bibitem{JML2010} J. M. Laget, Phys. Lett. B \textbf{685}, 146 (2010).
\end{thebibliography}
